\documentclass[longauth]{aa}
\usepackage{lipsum}
\usepackage{comment}
\usepackage{placeins}
\usepackage{float}
\usepackage{graphicx}
\usepackage{siunitx}
\usepackage{orcidlink}
\usepackage[utf8]{inputenc}
\usepackage{booktabs}
\usepackage{multirow}
\usepackage{array}
\usepackage{subcaption}
\usepackage{hyperref}
\hypersetup{
  colorlinks=true,
  linkcolor=blue,
  citecolor=blue,
  urlcolor=blue
}

\AtBeginDocument{%
}
\AtBeginDocument{%
}
\AtBeginDocument{%
}

\AtBeginDocument{%
}
\usepackage{txfonts}

\begin{document} 

   \title{\texttt{CoLoRe-2LPT}: Lyman-$\alpha$ mock catalogues for the validation of DESI cosmological analyses}

\author{
M.~F.~Ruiz-Herrera Bernal\orcidlink{0009-0000-5572-6157}\thanks{ \email{manuelfrancisco.ruiz-herrera@ciemat.es}} \inst{1},
S.~Avila\orcidlink{0000-0001-5043-3662} \inst{1},
A.~Font-Ribera\orcidlink{0000-0002-3033-7312} \inst{2,3},
H.~K.~Herrera-Alcantar\orcidlink{0000-0002-9136-9609} \inst{4,5},
D.~Alonso \inst{6},
A.~Cuceu\orcidlink{0000-0002-2169-0595} \inst{7},
L.~Casas \inst{3},
F.~Sinigaglia\orcidlink{0000-0002-0639-8043} \inst{8,9},
J.~Aguilar \inst{7},
S.~Ahlen\orcidlink{0000-0001-6098-7247} \inst{10},
O.~Alves \inst{11},
U.~Andrade\orcidlink{0000-0002-4118-8236} \inst{12,11},
E.~Armengaud\orcidlink{0000-0001-7600-5148} \inst{5},
A.~Bault\orcidlink{0000-0002-9964-1005} \inst{7},
F.~Beutler\orcidlink{0000-0003-0467-5438} \inst{13},
D.~Bianchi\orcidlink{0000-0001-9712-0006} \inst{14,15},
M.~Bonici \inst{16},
A.~Brodzeller\orcidlink{0000-0002-8934-0954} \inst{7},
D.~Brooks \inst{17},
A.~Carnero Rosell\orcidlink{0000-0003-3044-5150} \inst{8,9},
J.~Chaves-Montero\orcidlink{0000-0002-9553-4261} \inst{3},
Z.~Chen \inst{13},
Y.~Cho \inst{18},
T.~Claybaugh \inst{7},
K.~S.~Dawson\orcidlink{0000-0002-0553-3805} \inst{19},
A.~de la Macorra\orcidlink{0000-0002-1769-1640} \inst{20},
A.~Dey\orcidlink{0000-0002-4928-4003} \inst{21},
W.~Elbers\orcidlink{0000-0002-2207-6108} \inst{22},
S.~Ferraro\orcidlink{0000-0003-4992-7854} \inst{7,23},
J.~E.~Forero-Romero\orcidlink{0000-0002-2890-3725} \inst{24,25},
G.~Gambardella \inst{26},
S.~{Gontcho A Gontcho}\orcidlink{0000-0003-3142-233X} \inst{27},
D.~Gonzalez\orcidlink{0009-0009-6485-640X} \inst{28},
A.~X.~Gonzalez-Morales\orcidlink{0000-0003-4089-6924} \inst{28},
C.~Gordon\orcidlink{0000-0003-2561-5733} \inst{17},
R.~Gsponer\orcidlink{0000-0002-7540-7601} \inst{29},
G.~Gutierrez \inst{30},
J.~Guy\orcidlink{0000-0001-9822-6793} \inst{7},
C.~Hahn\orcidlink{0000-0003-1197-0902} \inst{31},
M.~Herbold\orcidlink{0009-0000-8112-765X} \inst{32},
K.~Honscheid\orcidlink{0000-0002-6550-2023} \inst{33,34,32},
M.~Ishak\orcidlink{0000-0002-6024-466X} \inst{35},
S.~Jos \inst{10},
S.~Juneau\orcidlink{0000-0002-0000-2394} \inst{21},
N.~V.~Kamble\orcidlink{0009-0008-6707-2777} \inst{35},
T.~Karim\orcidlink{0000-0002-5652-8870} \inst{36,37},
D.~Kirkby\orcidlink{0000-0002-8828-5463} \inst{38},
A.~Kremin\orcidlink{0000-0001-6356-7424} \inst{7},
M.~Landriau\orcidlink{0000-0003-1838-8528} \inst{7},
A.~Leauthaud\orcidlink{0000-0002-3677-3617} \inst{39,40},
Q.~Li\orcidlink{0000-0003-3616-6486} \inst{19},
W.~Liu\orcidlink{0000-0002-6673-3106} \inst{41},
K.~Lodha\orcidlink{0009-0004-2558-5655} \inst{18,42},
O.~Manasoiu \inst{17},
M.~Manera\orcidlink{0000-0003-4962-8934} \inst{43,3},
P.~Martini\orcidlink{0000-0002-4279-4182} \inst{33,44,32},
A.~Meisner\orcidlink{0000-0002-1125-7384} \inst{21},
R.~Miquel \inst{2,3},
J.~Morawetz \inst{45},
P.~Mukherjee\orcidlink{0000-0002-2701-5654} \inst{18},
A.~Mu\~{n}oz-Guti\'{e}rrez \inst{20},
S.~Nadathur\orcidlink{0000-0001-9070-3102} \inst{46},
G.~Niz\orcidlink{0000-0002-1544-8946} \inst{28,47},
H.~E.~Noriega\orcidlink{0000-0002-3397-3998} \inst{48,20},
E.~Paillas\orcidlink{0000-0002-4637-2868} \inst{49,50},
N.~Palanque-Delabrouille\orcidlink{0000-0003-3188-784X} \inst{5,7},
J.~Pan\orcidlink{0000-0001-9685-5756} \inst{11},
M.~Pellejero Ibanez\orcidlink{0000-0003-4680-7275} \inst{13},
W.~J.~Percival\orcidlink{0000-0002-0644-5727} \inst{45,16,51},
C.~Poppett \inst{7,52,23},
F.~Prada\orcidlink{0000-0001-7145-8674} \inst{53},
H.~Pulido-Hern{\'a}ndez\orcidlink{0009-0009-7807-9218} \inst{54},
A.~P\'{e}rez-Fern\'{a}ndez\orcidlink{0009-0006-1331-4035} \inst{55},
I.~P\'erez-R\`afols\orcidlink{0000-0001-6979-0125} \inst{56},
C.~Ravoux\orcidlink{0000-0002-3500-6635} \inst{57},
J.~Rohlf\orcidlink{0000-0001-6423-9799} \inst{10},
G.~Rossi \inst{58},
R.~Ruggeri\orcidlink{0000-0002-0394-0896} \inst{59},
L.~Samushia\orcidlink{0000-0002-1609-5687} \inst{60,61},
E.~Sanchez\orcidlink{0000-0002-9646-8198} \inst{1},
C.~Saulder\orcidlink{0000-0002-0408-5633} \inst{55},
D.~Schlegel \inst{7},
H.~Seo\orcidlink{0000-0002-6588-3508} \inst{41},
J.~Silber\orcidlink{0000-0002-3461-0320} \inst{7},
T.~Simon\orcidlink{0000-0001-7858-6441} \inst{62},
M.~Siudek\orcidlink{0000-0002-2949-2155} \inst{26,9},
G.~Tarl\'{e}\orcidlink{0000-0003-1704-0781} \inst{11},
W.~Turner\orcidlink{0009-0008-3418-5599} \inst{33,44,32},
R.~Vaisakh\orcidlink{0009-0001-2732-8431} \inst{63},
M.~Vargas-Maga\~na\orcidlink{0000-0003-3841-1836} \inst{20},
B.~A.~Weaver \inst{21},
H.~Yang \inst{13},
and H.~Zhang\orcidlink{0000-0001-6847-5254} \inst{45,51}
}

\institute{
CIEMAT, Avenida Complutense 40, E-28040 Madrid, Spain 
\and Instituci\'{o} Catalana de Recerca i Estudis Avan\c{c}ats, Passeig de Llu\'{\i}s Companys, 23, 08010 Barcelona, Spain 
\and Institut de F\'{i}sica d'Altes Energies (IFAE), The Barcelona Institute of Science and Technology, Edifici Cn, Campus UAB, 08193, Bellaterra (Barcelona), Spain 
\and Institut d'Astrophysique de Paris. 98 bis boulevard Arago. 75014 Paris, France 
\and IRFU, CEA, Universit\'{e} Paris-Saclay, F-91191 Gif-sur-Yvette, France 
\and Department of Physics, University of Oxford, Denys Wilkinson Building, Keble Road, Oxford OX1 3RH, United Kingdom 
\and Lawrence Berkeley National Laboratory, 1 Cyclotron Road, Berkeley, CA 94720, USA 
\and Departamento de Astrof\'{\i}sica, Universidad de La Laguna (ULL), E-38206, La Laguna, Tenerife, Spain 
\and Instituto de Astrof\'{\i}sica de Canarias, C/ V\'{\i}a L\'{a}ctea, s/n, E-38205 La Laguna, Tenerife, Spain 
\and Department of Physics, Boston University, 590 Commonwealth Avenue, Boston, MA 02215 USA 
\and University of Michigan, 500 S. State Street, Ann Arbor, MI 48109, USA 
\and Leinweber Center for Theoretical Physics, University of Michigan, 450 Church Street, Ann Arbor, Michigan 48109-1040, USA 
\and Institute for Astronomy, University of Edinburgh, Royal Observatory, Blackford Hill, Edinburgh EH9 3HJ, UK 
\and Dipartimento di Fisica ``Aldo Pontremoli'', Universit\`a degli Studi di Milano, Via Celoria 16, I-20133 Milano, Italy 
\and INAF-Osservatorio Astronomico di Brera, Via Brera 28, 20122 Milano, Italy 
\and Perimeter Institute for Theoretical Physics, 31 Caroline St. North, Waterloo, ON N2L 2Y5, Canada 
\and Department of Physics \& Astronomy, University College London, Gower Street, London, WC1E 6BT, UK 
\and Korea Astronomy and Space Science Institute, 776, Daedeokdae-ro, Yuseong-gu, Daejeon 34055, Republic of Korea 
\and Department of Physics and Astronomy, The University of Utah, 115 South 1400 East, Salt Lake City, UT 84112, USA 
\and Instituto de F\'{\i}sica, Universidad Nacional Aut\'{o}noma de M\'{e}xico,  Circuito de la Investigaci\'{o}n Cient\'{\i}fica, Ciudad Universitaria, Cd. de M\'{e}xico  C.~P.~04510,  M\'{e}xico 
\and NSF NOIRLab, 950 N. Cherry Ave., Tucson, AZ 85719, USA 
\and Institute for Computational Cosmology, Department of Physics, Durham University, South Road, Durham DH1 3LE, UK 
\and University of California, Berkeley, 110 Sproul Hall \#5800 Berkeley, CA 94720, USA 
\and Departamento de F\'isica, Universidad de los Andes, Cra. 1 No. 18A-10, Edificio Ip, CP 111711, Bogot\'a, Colombia 
\and Observatorio Astron\'omico, Universidad de los Andes, Cra. 1 No. 18A-10, Edificio H, CP 111711 Bogot\'a, Colombia 
\and Institute of Space Sciences, ICE-CSIC, Campus UAB, Carrer de Can Magrans s/n, 08913 Bellaterra, Barcelona, Spain 
\and University of Virginia, Department of Astronomy, Charlottesville, VA 22904, USA 
\and Departamento de F\'{\i}sica, DCI-Campus Le\'{o}n, Universidad de Guanajuato, Loma del Bosque 103, Le\'{o}n, Guanajuato C.~P.~37150, M\'{e}xico 
\and Institute of Physics, Laboratory of Astrophysics, \'{E}cole Polytechnique F\'{e}d\'{e}rale de Lausanne (EPFL), Observatoire de Sauverny, Chemin Pegasi 51, CH-1290 Versoix, Switzerland 
\and Fermi National Accelerator Laboratory, PO Box 500, Batavia, IL 60510, USA 
\and Department of Astronomy, University of Texas at Austin, 2515 Speedway, TX 78712, USA 
\and The Ohio State University, Columbus, 43210 OH, USA 
\and Center for Cosmology and AstroParticle Physics, The Ohio State University, 191 West Woodruff Avenue, Columbus, OH 43210, USA 
\and Department of Physics, The Ohio State University, 191 West Woodruff Avenue, Columbus, OH 43210, USA 
\and Department of Physics, The University of Texas at Dallas, 800 W. Campbell Rd., Richardson, TX 75080, USA 
\and Center for Astrophysics $|$ Harvard \& Smithsonian, 60 Garden Street, Cambridge, MA 02138, USA 
\and Department of Astronomy \& Astrophysics, University of Toronto, Toronto, ON M5S 3H4, Canada 
\and Department of Physics and Astronomy, University of California, Irvine, 92697, USA 
\and Department of Astronomy and Astrophysics, UCO/Lick Observatory, University of California, 1156 High Street, Santa Cruz, CA 95064, USA 
\and Department of Astronomy and Astrophysics, University of California, Santa Cruz, 1156 High Street, Santa Cruz, CA 95065, USA 
\and Department of Physics \& Astronomy, Ohio University, 139 University Terrace, Athens, OH 45701, USA 
\and University of Science and Technology, 217 Gajeong-ro, Yuseong-gu, Daejeon 34113, Republic of Korea 
\and Departament de F\'{i}sica, Serra H\'{u}nter, Universitat Aut\`{o}noma de Barcelona, 08193 Bellaterra (Barcelona), Spain 
\and Department of Astronomy, The Ohio State University, 4055 McPherson Laboratory, 140 W 18th Avenue, Columbus, OH 43210, USA 
\and Department of Physics and Astronomy, University of Waterloo, 200 University Ave W, Waterloo, ON N2L 3G1, Canada 
\and Institute of Cosmology and Gravitation, University of Portsmouth, Dennis Sciama Building, Portsmouth, PO1 3FX, UK 
\and Instituto Avanzado de Cosmolog\'{\i}a A.~C., San Marcos 11 - Atenas 202. Magdalena Contreras. Ciudad de M\'{e}xico C.~P.~10720, M\'{e}xico 
\and Instituto de Ciencias F\'{\i}sicas, Universidad Nacional Aut\'onoma de M\'exico, Av. Universidad s/n, Cuernavaca, Morelos, C.~P.~62210, M\'exico 
\and Instituto de Estudios Astrof\'isicos, Facultad de Ingenier\'ia y Ciencias, Universidad Diego Portales, Av. Ej\'ercito Libertador 441, Santiago, Chile 
\and Steward Observatory, University of Arizona, 933 N. Cherry Avenue, Tucson, AZ 85721, USA 
\and Waterloo Centre for Astrophysics, University of Waterloo, 200 University Ave W, Waterloo, ON N2L 3G1, Canada 
\and Space Sciences Laboratory, University of California, Berkeley, 7 Gauss Way, Berkeley, CA  94720, USA 
\and Instituto de Astrof\'{i}sica de Andaluc\'{i}a (CSIC), Glorieta de la Astronom\'{i}a, s/n, E-18008 Granada, Spain 
\and Departamento de F\'{i}sica, Instituto Nacional de Investigaciones Nucleares, Carreterra M\'{e}xico-Toluca S/N, La Marquesa,  Ocoyoacac, Edo. de M\'{e}xico C.~P.~52750,  M\'{e}xico 
\and Max Planck Institute for Extraterrestrial Physics, Gie\ss enbachstra\ss e 1, 85748 Garching, Germany 
\and Departament de F\'isica, EEBE, Universitat Polit\`ecnica de Catalunya, c/Eduard Maristany 10, 08930 Barcelona, Spain 
\and Universit\'{e} Clermont-Auvergne, CNRS, LPCA, 63000 Clermont-Ferrand, France 
\and Department of Physics and Astronomy, Sejong University, 209 Neungdong-ro, Gwangjin-gu, Seoul 05006, Republic of Korea 
\and Queensland University of Technology,  School of Chemistry \& Physics, George St, Brisbane 4001, Australia 
\and Abastumani Astrophysical Observatory, Tbilisi, GE-0179, Georgia 
\and Department of Physics, Kansas State University, 116 Cardwell Hall, Manhattan, KS 66506, USA 
\and Sorbonne Universit\'{e}, CNRS/IN2P3, Laboratoire de Physique Nucl\'{e}aire et de Hautes Energies (LPNHE), FR-75005 Paris, France 
\and Department of Physics, Southern Methodist University, 3215 Daniel Avenue, Dallas, TX 75275, USA
}

\abstract{The Lyman-$\alpha$ (Ly$\alpha$) forest has become a crucial probe for studying the large-scale structure of the universe at high redshift ($z > 2$), providing powerful constraints on Baryon Acoustic Oscillations (BAO) and the full-shape (FS) clustering of matter. As a key ingredient for upcoming BAO and FS analyses, we present a new generation of fast cosmological Ly$\alpha$ mocks based on second-order Lagrangian perturbation theory (2LPT). These new mocks significantly improve upon previous log-normal approaches, both at accurately capturing small scale clustering and at recovering the non-linear broadening of the BAO peak. They are able to reproduce Ly$\alpha$ statistics within $10\%$ of the latest DESI measurement; including the Ly$\alpha$ bias and the redshift-space distortion $\beta$ parameter, mean transmitted flux, and 1D power spectrum. The corresponding quasar (QSO) clustering is also improved with respect to previous approaches, calibrated against high-resolution \texttt{Abacus} simulations, recovering the observational QSO linear bias to less than $5\%$ and improving redshift-space distortions via 2LPT velocities and the addition of Fingers-of-God effects. Furthermore, these mocks incorporate high column density systems and metal lines, allowing us to explore the effects and systematics induced by these astrophysical contaminants. This new set of mocks has been key for enhancing the modeling and validation of the DESI DR2 Ly$\alpha$ full shape cosmological analysis. This work provides a physically motivated and computationally efficient tool for simulating current and next-generation Ly$\alpha$ surveys and validating FS and BAO analysis.}

\titlerunning{\texttt{CoLoRe-2LPT}: Lyman-$\alpha$ mock catalogues for the validation of DESI cosmological analysis.}
\authorrunning{Ruiz-Herrera Bernal et al.}
\maketitle

\section{Introduction}
Obtaining precise measurements of the expansion history of the universe and understanding its acceleration remain a central goal of modern cosmology. Although the standard $\Lambda$CDM model provides a successful description of many observations, recent high-precision datasets have begun to reveal tensions and possible indications that dark energy may not be fully described by a simple cosmological constant \citep{DESBAOSN, DESIII}. Progress in this area requires not only increasingly accurate measurements of cosmological distances but also robust constraints on the growth rate of large-scale structure.

The Lyman-$\alpha$ (Ly$\alpha$) forest has emerged as a uniquely powerful probe for studying the large-scale structure of the high-redshift ($z > 2$) universe. Produced by the absorption of neutral hydrogen in the intergalactic medium along the lines of sight (LOS) to distant quasars (QSOs), the Ly$\alpha$ forest continuously traces the matter density field  in three dimensions at redshifts $2\leq z \leq 4$ \citep{Bi,Croft,Hui}. This makes it complementary to galaxy surveys, which primarily probe lower redshifts and provide discrete tracers of structure. Early Ly$\alpha$ analyses focused primarily on measuring the Baryon Acoustic Oscillation (BAO) scale, first detected using data from the Baryon Oscillation Spectroscopic Survey (BOSS) \citep{Dawson2013} via the Ly$\alpha$ auto-correlation \citep{SlosarBAO, BuscaBAO, Kirkby2013, Delubac15, Bautista}, and also the Ly$\alpha \times\rm{QSO}$ cross-correlation \citep{AndreuQSOs, DMDB17}. These measurements have subsequently been improved with later datasets like extended BOSS (eBOSS) \citep{Dawson, SainteAgathe, Blomqvist2019, eBOSSBAO}.

However, beyond BAO, the Ly$\alpha$ forest contains substantially more large-scale cosmological information encoded in the full-shape (FS) of its three-dimensional correlation functions \citep{CuceuFS1, CuceuFS2}. In particular, the anisotropic structure of Ly$\alpha$ correlations provides access to two key effects. First, the Alcock–Paczynski (AP) \citep{AP79} effect introduces geometric distortions when an incorrect fiducial cosmology is assumed, enabling direct constraints on the expansion rate and angular diameter distance. Second, redshift-space distortions (RSD) arise from peculiar velocities and probe the growth rate of cosmic structure. Together with BAO, these effects capture most of the large-scale cosmological information available in Ly$\alpha$ clustering.

The Dark Energy Spectroscopic Instrument (DESI) \citep{DESIinstr2} is a wide-field spectroscopic survey mounted on the Mayall 4-m telescope designed to map the large-scale structure of the universe with unprecedented statistical power. Its focal plane hosts thousands of robotic fibre positioners \citep{FocalPlane, FiberSystem} feeding ten spectrographs through a wide-field optical corrector \citep{Corrector}. The survey operations and data reduction pipeline \citep{SurveyOps, GuyPipeline} transform the raw observations into analysis-ready catalogues, enabling the acquisition of spectra for hundreds of thousands of high-redshift quasars \citep{DESIgeneral, DESIDR1}. This unprecedented increase in data is ushering BAO \citep{DESII, DESIII} and FS analyses \citep{CuceuFS0, CuceuFS1, CuceuFS2, DESIFSDR1} into a new era of precision, enabling the most stringent cosmological constraints to date \citep{DESII, DESIII}. However, with this increased precision comes greater sensitivity to modeling assumptions, astrophysical systematics, and observational effects \citep{Guy2025}. Consequently, ensuring that BAO and FS analyses remain robust and unbiased demands thorough and extensive validation.

Realistic mock catalogues (mocks) are an essential component of this validation process. They enable testing of analysis pipelines, quantification of systematic biases, evaluation of covariance matrices, and verification of parameter recovery under controlled conditions \citep{CuceuMock, LauraCasas}. For Ly$\alpha$ forest studies, such mocks must reproduce the forest and QSO large-scale clustering together with a realistic representation of the astrophysical contaminants present in the data and also the survey geometry and observational effects. Additionally, these mocks must be efficient enough to generate large statistical ensembles covering the survey volume and Ly$\alpha$ redshift range -- something infeasible for current N-body or hydrodynamical simulations \citep[e.g.][]{Nyx2, Nyx1, Boryana}.

In previous analyses \citep[e.g.][]{eBOSSBAO, DESIII, CuceuFS2}, cosmological Ly$\alpha$ mocks were typically generated from log-normal realizations of the density field \citep{Coles91, CoLoRe, Saclay}, which were then post-processed to produce Ly$\alpha$ transmission skewers \citep{LyaCoLoRe}. While these mocks can approximately reproduce clustering on large scales, such as those relevant for BAO, they are unreliable for studying mildly non-linear scales i.e., intermediate scales ($k\sim0.1 - 0.5$ $h$/Mpc, or separations of roughly $10 - 80$ Mpc/$h$). These limitations were explored in \citet{LauraCasas} for DESI DR2 Ly$\alpha$ BAO validation, where improvements were made to the QSO clustering and an approximate BAO broadening was introduced by modifying the input power spectrum. Nevertheless, the small-scale features in these mocks remain governed by the log-normal model, rendering them insufficiently realistic for FS analyses, particularly in the Ly$\alpha$ case, where the scale mixing effect of the distortion matrix needs to be taken into account.

In this work, we present a new suite of Ly$\alpha$ forest mock catalogues designed specifically to validate the DESI DR2 FS Ly$\alpha$ analysis, while also serving future BAO and FS studies. The key novelty of these mocks is the use of second-order Lagrangian perturbation theory (2LPT) \citep{Moutarde1991, Bouchet1995} to evolve the density field. This approach naturally captures mild non-linearities relevant for FS analysis and includes BAO broadening. In addition, we also improve the QSO clustering by adopting biasing models calibrated against \texttt{Abacus} N-body simulations \citep{AbacusSummit, AbacusHOD}, and generate Ly$\alpha$ forest skewers with clustering statistics that reproduce the latest DESI measurements \citep{Ravoux, Turner, Karacayli25, DESIII}. Despite this increased realism, these simulations retain computational efficiency, enabling the generation of large ensembles of realizations. 

Although the usage of LPT methods \citep{PTHalos, Pinocho, COLA,PATCHY, Halogen,EZMocks} has been widely explored in the production of mocks for galaxy clustering surveys \citep{Manera2LPT, Kitaura2016, Avila2018, Ferrero2021,Zhao2021, EuclidSkies}, this work constitutes the first time they are applied in the context of full light cone mocks for a Ly$\alpha$ forest survey.\footnote{Note that the LPT scheme has been previously used for Ly$\alpha$ mocks but limited to boxes in \citet{Francesco1}. Efforts to extend the ALPT framework of \citet{Francesco1} to light cones are ongoing (Sinigaglia et al., in prep).} The implementation we used here is based in the public codes \texttt{CoLoRe} \citep{CoLoRe}, which is designed to run lightcones with multiple cosmological observables simultaneously, and \texttt{LyaCoLoRe} \citep{LyaCoLoRe}, which takes density and velocity skewers from \texttt{CoLoRe} and transforms them into Ly$\alpha$ transmission skewers.

The primary goal of this paper is to describe the techniques used to build these new mocks and their validation, while their application to the DESI DR2 Ly$\alpha$ FS analysis will be presented in the companion publication \citep{DESI:2026DR2IV}. 

The paper is organized as follows. In \autoref{sec:colore}, we describe and validate our methodology for generating 2LPT density fields and populating them with QSOs to achieve clustering consistent with $N$-body simulations. In \autoref{sec:lyacolore}, we detail the post-processing of these density fields into Ly$\alpha$ transmission skewers with the correct clustering and statistical properties, which we validate in \autoref{sec:lya_val}. \autoref{sec:gen_desi} describes the incorporation of astrophysical contaminants and the generation of DESI-like mocks. In \autoref{sec:val_cont}, we validate these contaminated DESI-like mocks by comparing them directly to DESI DR2 BAO results. Finally, \autoref{sec:conclusions} summarizes the properties of these mocks and their potential applications for current and future studies.

Throughout the paper, we will use the mean estimates of the \texttt{TT,TE,EE+lowE+lensing} likelihood chains from the Planck 2018 results \citep{PLII} as our fiducial cosmology.

\section{\texttt{CoLoRe-2LPT}: Creating cosmological light cones with QSOs}\label{sec:colore}
The starting point of our mocks is the code \texttt{CoLoRe}\footnote{\url{https://github.com/damonge/CoLoRe/tree/master}.}\citep{CoLoRe}. \texttt{CoLoRe} is a highly parallelized code that generates large cosmological light cones with multiple tracers following the same matter density field. For our purposes, we will use \texttt{CoLoRe} to populate light cones with QSOs and get the line of sight (LOS) from each of them to the observer, which will be the seed of our Ly$\alpha$ transmission skewers.

Previous \texttt{CoLoRe} Ly$\alpha$ mocks \citep{LyaCoLoRe, LauraCasas} relied completely on the fact that a Gaussian field is transformed into density via a log-normal transformation \citep{Coles91}. While the original version of {\tt CoLoRe} also allowed for the use of first- and second-order LPT to generate the matter density field, this implementation was not complete, as it had not been fully validated, and lacked the ability to simulate the velocity field at the same LPT order. Here, we present \texttt{CoLoRe-2LPT}\footnote{Publicly available at \url{https://github.com/damonge/CoLoRe/tree/master}.}, a new version of \texttt{CoLoRe} that addresses this shortcoming, allowing for the use of 2LPT in the simulation of all cosmological tracers. This approach aims to improve non-linearities and small scale clustering while maintaining its low computational requirements to the greatest extent possible to generate hundreds of realizations. 

The \texttt{CoLoRe-2LPT} pipeline can be summarized as follows:
\begin{enumerate}
    \item Create a box at $z=0$ with fixed resolution and size (given by the maximum redshift of the light cone), and populate each cell with a Gaussian random field according to an input power spectrum and cosmology.
    \item Evolve a set of test particles using 2LPT with the Gaussian random field as input and get their first- and second-order LPT displacements at $z=0$. 
    \item Generate a light cone by rescaling the particles positions and velocities via growth factor and rate depending on the redshift of each cell with respect to the observer (center of the box).
    \item Convert the particles positions and velocities into smooth density and velocity fields by interpolating them into a grid.
    \item Populate the light cone with QSOs by:
    \begin{itemize}
        \item Applying a biasing model to the density field to reproduce the observed QSO clustering.
        \item Randomly sample  biased cells with uniform distribution to match the input QSO number density. 
    \end{itemize} 
    \item Perform a trilinear interpolation of the 2LPT density and velocity fields to get the LOS skewers from each QSO to the observer.
\end{enumerate}

The main differences from previous versions arise in steps 2-5, while steps 1 and 6 remain largely unchanged up to minor numerical corrections. We describe steps 2–4 in detail, together with their validation, in \autoref{sec:2lpt} and \autoref{sec:2lpt_val}. In \autoref{sec:qso_clustering}, we present the QSO biasing model corresponding to step 5 and calibrate it using high-resolution \texttt{Abacus} simulations.

\subsection{Lagrangian Pertubation Theory}\label{sec:2lpt}
The basis of 2LPT -- and Lagrangian perturbation theory in general -- is to perform an $n^{\rm{th}}$-order perturbation of the displacement field of a set of particles based on linear over-densities -- in our case, the initial Gaussian random field. The final non-linear density field is then defined by the spatial distribution of these displaced particles. For a complete description of this theory, see \citet{Moutarde1991, Bouchet1995, 2LPTIc}. 

In this sense, we start with massive test particles in some fixed positions -- in our case, the center of the \texttt{CoLoRe-2LPT} grid cells -- and perturb them so that the final physical position of each particle at time $t$ is given by: 
\begin{equation}
    x(t) = a(t)\Big[q + \Psi(q,t)\Big],
\end{equation}
where $q$ is the initial position in comoving coordinates, $a(t)$ is the scale factor and $\Psi(q,t)$ is the 2LPT displacement vector, which is given by the linear matter overdensity. 

The relation between the 2LPT displacement vector and the initial Gaussian field is given by Newton's second law subject to the particles gravitational potential. The 2LPT curl-free solution of this equation can be written as:
\begin{equation}\label{eq:2lptsol}
\boldsymbol{\Psi}(\mathbf{q}, a) =
D_{1}(a)\,\nabla \phi_{1}(\mathbf{q})
+ D_{2}(a)\,\nabla \phi_{2}(\mathbf{q}),
\end{equation}
where, for \texttt{CoLoRe-2LPT}, $\phi_{1}$ and $\phi_{2}$ are the 1LPT and 2LPT potentials solving the following system of equations for the initial Gaussian random field $\delta_{G}$ in Fourier space:
\begin{equation}
\left\{
\begin{aligned}
\nabla^{2} \phi_{1} &= -\delta_{G} , \\
\nabla^{2} \phi_{2} &=
\frac{1}{2}
\sum_{ij}
\partial_i^{2} \phi_{1}\,
\partial_j^{2} \phi_{1}
-
\left(
\partial_i \partial_j \phi_{1}
\right)^{2},
\end{aligned}
\right.
\end{equation}
and $D_{1}$ and $D_{2}$ are the first- and second-order growth factors. In \texttt{CoLoRe-2LPT} we compute $D_{1}$ from its complete definition and for $D_{2}$ we use the approximation \citep{Bouchet1995}:
$D_{2}(a) =
-\frac{3}{7}
\,D_{1}^{2}(a)
\,\Omega_{\mathrm{M}}^{-1/143}(a)
$ .
Finally, once we have the physical position of each particle, their physical velocity will be:
\begin{equation}
v(t) \equiv \frac{dx(t)}{dt} = H(a)\Big(D_{1}(a)f_{1}(a)\,\nabla \phi_{1}(\mathbf{q})+ D_{2}(a)f_{2}(a)\,\nabla \phi_{2}(\mathbf{q})\Big),
\end{equation}
where $H$ is the Hubble parameter and $f_{1}$ and $f_{2}$ are the first and second order growth rates. 
Consistently with the second order growth factor approximation, we solve completely for $f_{1}$ and approximate $f_{2}$ as:
$
f_{2}(a) = 2\Omega_{M}^{6/11}(a)$ \citep{Bouchet1995}.

Note that the LPT potentials will be computed at $z=0$ and then rescaled by the growth factors and rates according to the redshift of each cell with the observer as in \autoref{eq:2lptsol}. Finally, the non-linear 2LPT density and velocity fields result from the interpolation of these particles properties onto the box grid (see the end of next section).

\subsection{Validation of the 2LPT fields}\label{sec:2lpt_val}
In this section, we validate the implementation of 2LPT at the particle and density field level in real and redshift space. We start by comparing particle clustering from the snapshot version of \texttt{CoLoRe-2LPT}\footnote{Publicly available at \url{https://github.com/damonge/CoLoRe/tree/snap}.} to  \texttt{2LPTic}\footnote{\url{https://cosmo.nyu.edu/roman/2LPT/}.}\citep{2LPTIc}, a code widely tested in the literature that only creates dark matter snapshots.

We generate 50 realizations of both codes with box size $L_{\rm{box}} = 1$ Gpc/h and a particle and grid number of $N_{\rm{particles}} = n_{\rm{grid}} = 512^{3}$ for $z \in \{0,1,2,...,10\}$. We then compute the power spectrum and two point correlation function (2PCF) multipoles using the public \texttt{python} package \texttt{nbodykit}\footnote{\url{https://github.com/bccp/nbodykit}.}.

\begin{figure}[h]
    \centering
    \includegraphics[width=\linewidth]{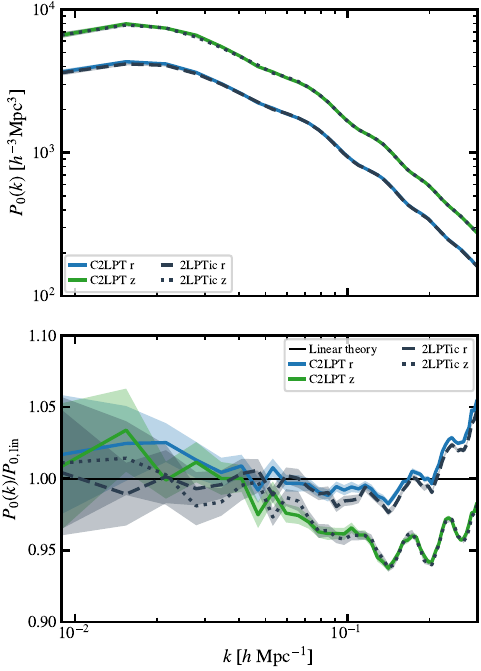}
    \caption{\textit{Top}: \texttt{CoLoRe-2LPT} (C2LPT) particle power spectrum monopole in real space (r) and redshift space (z) compared to \texttt{2LPTic} particles at z=2 snapshot. \textit{Bottom}: Ratio of these monopoles to linear theory. In both panels, the lines for each code represent the mean of 50 realizations in boxes with size $L_{\rm{box}}=1$ Gpc/h and $512^{3}$ particles. Shades represent the error on the ensemble average ($\sigma_{\rm{mean}}$) for these 50 realizations.}
    \label{Figure1}
\end{figure}

In the top panel of \autoref{Figure1} we show the comparison for the mean power spectrum monopole in real space at $z=2$ over the 50 realizations and its error on the ensemble average ($\sigma_{\rm{mean}}= \sigma/\sqrt{N_{\rm{mocks}}}$, $N_{\rm{mocks}}$ being the number of realizations and $\sigma$ the standard deviation). In the bottom panel, we present their ratio to linear theory. We find that the agreement between the codes is consistent within uncertainties\footnote{Note that both sets of mocks use different initial seeds and thus we don't expect the curves to be exactly the same.}, with both exhibiting a comparable level of non-linearity at high $k$ relative to linear theory. 

\begin{figure}[h]
    \centering
    \includegraphics[width=\linewidth]{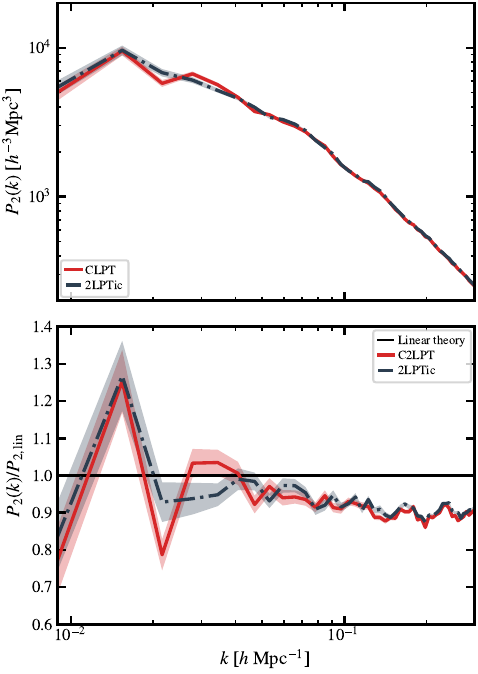}
    \caption{\textit{Top}: \texttt{CoLoRe-2LPT} (C2LPT) particles power spectrum quadrupole in redshift space compared to \texttt{2LPTic} particles at a z=2 snapshot. \textit{Bottom}: Ratio of these quadrupoles to linear theory at the same redshift. In both panels, the lines for each code represent the mean of 50 realizations in boxes with size $L_{\rm{box}}=1$ Gpc/h and $512^{3}$ particles. Shades represent the error on the ensemble average ($\sigma_{\rm{mean}}$) for these 50 realizations.}
    \label{Figure2}
\end{figure}

For the redshift-space case, we just add the velocity displacement to the particles positions. We present the redshift-space results for the monopole and quadrupole in \autoref{Figure1} and \autoref{Figure2}, respectively. The hexadecapole is not shown since it is largely noise dominated. Similarly to real space, both codes agree within the statistical uncertainties, showing the same level of non-linearity. We also verified that all results are consistent for different snapshot redshifts, as well as for the configuration-space 2PCF.

The only potential concern could be the oscillatory pattern seen in the first $k$ bins for the quadrupole. However, as can be deduced from the similar behavior between the codes, this pattern is just a binning artifact that couples to the discreteness of the Fourier modes in the power spectrum computation. To verify this, we rerun these tests for boxes with $L_{\rm{box}}=2$ Gpc/h and $n_{\rm{grid}} = 1024^{3}$ and concluded that the problem alleviates for these $k$ bins and still shows the same behavior in both codes.

It is also worth stating that in all these tests we also checked the effect of smoothing the input Gaussian power spectrum of \texttt{CoLoRe} \citep[][section 2.3.1.]{CoLoRe}. This functionality was key for the log-normal version of \texttt{CoLoRe} since it allowed to control divergences in the density field due to the log-normal transformation of the Gaussian initial field \citep{LyaCoLoRe, LauraCasas}. However, for the \texttt{CoLoRe-2LPT} case we find that the effect of this smoothing is negligible and we decide to switch it off for the rest of our studies.

Finally, since \texttt{CoLoRe} works with fields on a regular grid, we convert the 2LPT particles positions and velocities into smooth fields using a cloud in cell (CIC) mass and momentum mesh interpolation scheme. For the momentum case, we set cells with zero density to zero velocity since we find that this have a negligible impact in our results. To validate this step, we populated the box with unity bias tracers (see \autoref{sec:qso_clustering}) and checked that their clustering is consistent with that of the particles in both real and redshift space up to mesh effects at the scale of the interpolation grid.

\subsection{QSO clustering}\label{sec:qso_clustering}
In this section, we aim to populate the 2LPT density field with QSOs with clustering properties matched to a reference \texttt{Abacus} simulation \citep{AbacusSummit, AbacusHOD}. 
The starting point is the local, density-dependent, simplified threshold bias model \citep{Kaiser2, BBKS}:
\begin{equation}\label{eq:biasing}
1 + \delta_Q =
\begin{cases}
1 + b_Q \, \delta, & \text{if } \delta > t, \\
0, & \text{if } \delta \leq t,
\end{cases}
\end{equation}
these equations represent a linear biasing model, where the density of quasars is proportional to the density of matter via $b_{Q}$, but modified so that quasars can only populate cells with a matter density above the threshold parameter $t$ -- see \citet{Francesco1, LauraCasas} for similar approaches.

In the case of log-normal quasi-linear \texttt{CoLoRe} realizations (hereafter \texttt{CoLoRe-QL} mocks) for DESI DR2 BAO, this model was shown to improve the QSO clustering from previous mocks, matching \texttt{Abacus} at intermediate scales \citep[][Fig 1.]{LauraCasas}. In the 2LPT case, unlike in log-normal/Gaussian realizations, the relation between the QSO bias ($b_{\rm{QSO}}$) and the biasing scheme parameters ($b_{Q}$ and $t$) is not analytically defined. Hence, we will find empirical relations.  

\subsubsection{The QSO power spectrum in snapshots.}\label{sec:qso_snap}
 We start by fitting $(b_{Q}, t)$ at a few redshifts using snapshots and then derive a redshift dependent model to implement in our light cones. 

We employ \texttt{Abacus} Halo Occupation Distribution (HOD) QSO snapshot catalogues \citep{AbacusHOD} matching the DESI DR1 QSO linear bias measurements \citep{ChaussidonQSO} to calibrate our biasing model. This approach allows us to work efficiently in Fourier space without the need to correct for the impact of a window function.

The model calibration is performed in a ($b_{Q},t$) grid for  $z=2.0, 2.5$ and $3.0$. We produce 25 \texttt{CoLoRe-2LPT} realizations with $L_{\rm{box}} = 2$ Gpc/h (same as \texttt{Abacus}), $n_{\rm{grid}} = 1024^{3}$ (yielding the same resolution as the final configuration, \autoref{sec:colorerun}) and the same number density of QSOs as \texttt{Abacus} to match the shot noise. We perform a $\chi^2$ minimization between the mean power spectra of \texttt{Abacus} and \texttt{CoLoRe-2LPT} considering the quadratic sum of the error on the ensemble average ($\sigma_{\rm{mean}}$) over all bins from the fundamental mode $k_f$ up to a maximum scale $k_{\max}$. This results mostly in matching the linear bias, but allows the possibility to fit other terms among the two simulations. This minimization is computed in real space for $k_{max} = 0.08, 0.09$ and $0.1$ $h$/Mpc.

We run this procedure first for a coarse mesh in the $(b_{Q}, t)$ space and then iteratively refined the mesh around the optimal values achieving a final resolution of $0.001$ for both parameters. We find that the parameter $t$ is well constrained across all redshifts and choices of $k_{\max}$. However, the parameter $b_{Q}$ varies to effectively tilt the power spectrum and improve agreement with \texttt{Abacus} on small scales. Motivated by this behavior, we fix $b_{Q}$ to its best-fit value obtained at $k_{\max} = 0.1$ $h$/Mpc. Our final best-fit $(b_{Q}, t)$ parameters are $(2.720, 0.985), (3.310,1.384)$ and $(4.020, 1.595)$ for redshift $z=2.0,2.5$ and $3.0$ respectively.

\begin{figure}[!htbp]
    \centering
    \includegraphics[width=\linewidth]{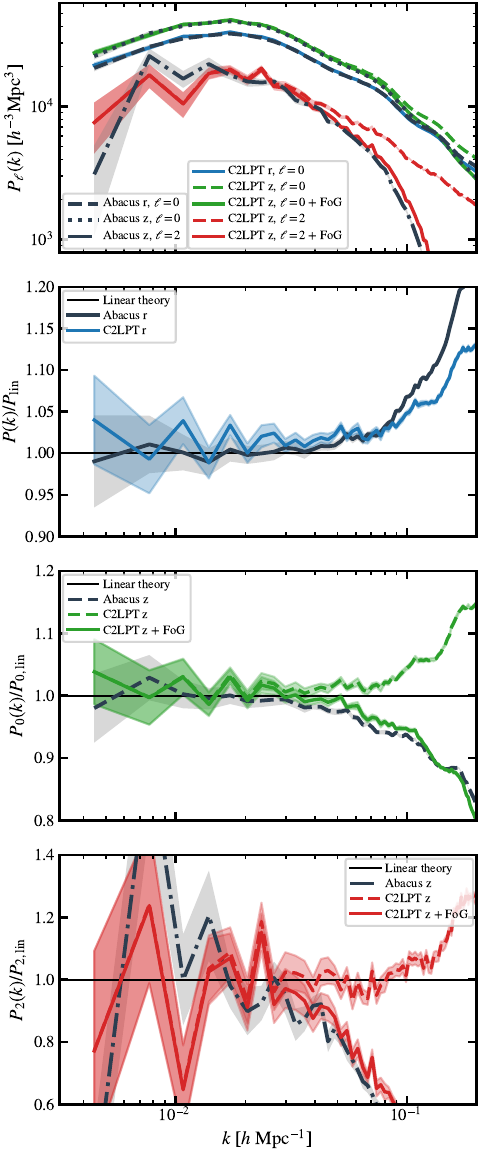}
    \caption{\textit{Top:} Comparison of \texttt{CoLoRe-2LPT} (C2LPT) best fit QSO power spectrum monopoles and quadrupole in real (r) and redshift (z) space against \texttt{Abacus} at z=2 snapshot. In the rest of the panels we show the ratio of each multipole to (RSD) linear theory. In all panels the lines represent the mean of 25 realizations of each code and the shade represents $\sigma_{\rm{mean}}$. For the \texttt{CoLoRe-2LPT} case, solid lines represent realizations including the Fingers-of-God effect.}
    \label{Figure3}
\end{figure}

In \autoref{Figure3} we show the \texttt{Abacus} and \texttt{CoLoRe-2LPT} mean power spectrum monopole in real space -- denoted with r -- and its $\sigma_{\rm{mean}}$ as a shade at $z=2$ for the best fit values for $k_{max}=0.1$ $h$/Mpc. In the second panel, we present the ratio of both \texttt{Abacus} and \texttt{CoLoRe-2LPT} to linear theory assuming the \texttt{Abacus} best fit bias \citep{AbacusHOD}. The \texttt{CoLoRe-2LPT} monopole is consistent with the \texttt{Abacus} one within $\sigma_{\rm{mean}}$ up to $k_{max}\sim0.1$ $h$/Mpc, which confirms that our fitting procedure works. Furthermore, the departure from linear theory is very similar for both simulations at large and intermediate $k$. For $k > k_{max}$ the \texttt{Abacus} monopole exceeds \texttt{CoLoRe-2LPT} in power as expected, given that its density field presents more realistic non-linearities as it is based on an N-body simulation.

We also present the mean \texttt{Abacus} and \texttt{CoLoRe-2LPT} redshift space monopole and quadrupole -- denoted with z -- at $z=2$ for the same best fit parameters in the top panel of \autoref{Figure3}. In the third and fourth panels we present the monopole and quadrupole ratio of the mocks to RSD linear theory \citep{Kaiser} as dashed lines. In this case, the agreement between \texttt{Abacus} and \texttt{CoLoRe-2LPT} is reduced to $k\sim0.03$ $h$/Mpc for the monopole and $k\sim0.02$ $h$/Mpc for the quadrupole. For $k$ higher than this, \texttt{CoLoRe-2LPT} shows an increase of power with respect to linear theory, in line with the real space behavior. However, \texttt{Abacus} shows a decrease of power at large $k$ due to the inclusion of Fingers-of-God (FoG) effect which produces a damping due to random QSO (galaxy) motions \citep{FoG}. 

In order to improve the smaller scales, we model this FoG effect in \texttt{CoLoRe-2LPT} by adding a random velocity to the QSOs coming from a normal distribution \citep{Hess2013} with dispersion of $\sigma_v =350$ km/s that roughly fits the clustering of \texttt{Abacus}; see solid lines in \autoref{Figure3}. 
The multipoles including FoG improve the agreement with \texttt{Abacus} up to smaller scales ($k\gtrsim0.1$ $h$/Mpc), presenting the typical damping with respect to linear theory.

\subsubsection{The QSO auto-correlation in light cones: Redshift dependent biasing scheme}\label{sec:qso_auto}
The $(b_{Q},t)$ values obtained with the previous procedure ensure that the QSO clustering in \texttt{CoLoRe-2LPT} snapshots is realistic when benchmarked against \texttt{Abacus} at $z=2.0, 2.5$ and $3$. However, the final goal is to obtain a biasing scheme with redshift dependency that can be applied directly to light cones.

We approximately extrapolate the information from these discrete points by assuming a redshift dependency for the $(b_{Q},t)$ parameters and check the resultant QSO bias against linear theory. For the $b_{Q}$ parameter, we assume a power law fit based on the redshift evolution of the observational QSO bias \citep{ChaussidonQSO}:
\begin{equation}
b_{Q}(z) = A \, (1+z)^{B}.
\end{equation}
where the parameters are: $A = 0.523$ and $B = 1.475$.

For the threshold parameter ($t$), inspired by the results obtained in the previous section, we propose the following redshift dependence:
\begin{equation}
t(z) = y_1 + \Delta
       \frac{1 - r^{\frac{z-2}{h}}}{1 - r},
\end{equation}
where $y_1$ is the threshold value at $z = 2$, $\Delta$ sets the increment per redshift step of width $h$, and $r$ is the geometric ratio controlling how quickly the growth of $t(z)$ decays. This equation represents an empirically motivated function that increases with redshift at a geometrically declining rate, asymptoting to $y_{1}+\Delta/(1-r)$, valid for the high redshifts of our analysis ($z>1.8$). The best-fit values are $y_1 = 0.984$, $r = 0.523$, $h = 0.5$, $\Delta = 0.401$.

\begin{figure}[h]
    \centering
    \includegraphics[width=\linewidth]{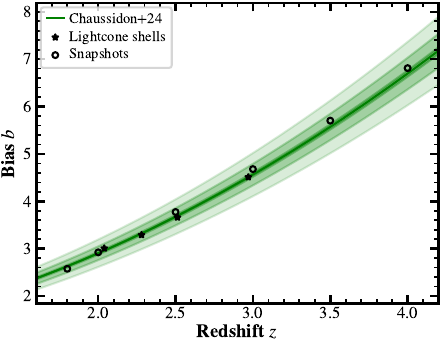}
    \caption{QSO bias in \texttt{CoLoRe-2LPT} snapshots and light cone shells compared to DESI DR1 results \citep{ChaussidonQSO} at different effective redshifts. We represent with dots the best linear fit bias for \texttt{CoLoRe-2LPT} snapshots QSO power spectra and with stars for light cone QSO auto-correlations. The shadows represent, with decreasing intensity, the 1, 5 and 10 \% of the DESI DR1 measurement.}
    \label{Figure4}
\end{figure}

We implement this new biasing-scheme directly on \texttt{CoLoRe-2LPT} and measure the power spectrum in real space for $z \in \{1.8, 2.0, 2.5, 3.0, 3.5, 4.0\}$ snapshots. We then fit these power spectra with linear theory leaving only the QSO bias as free parameter up to $k_{max}=0.1$ $h$/Mpc. We present the measured biases in \autoref{Figure4} compared with this latest DESI DR1 measurement of \citet{ChaussidonQSO} represented as a solid line and its $1\%$, $5\%$ and $10\%$\footnote{Here and throughout the rest of the paper, by a percentage $X\%$ of a measurement we mean the fractional deviation from its central value independent of its statistical uncertainty.} as decreasing intensity shades. We can see that the \texttt{CoLoRe-2LPT} bias is within $5\%$ or less for all redshifts.

\begin{figure}[h]
    \centering
    \includegraphics[width=\linewidth]{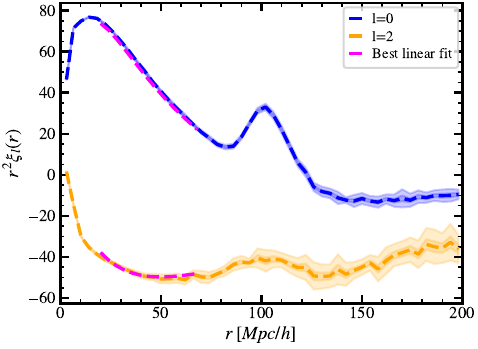}
    \caption{QSO 2PCF monopole and quadrupole for the stack of 10 full sky \texttt{CoLoRe-2LPT} light cones with $z_{\rm{eff}}=2.33$. The shadows represent $\sigma_{\rm{mean}}$ for these 10 realizations. The dashed magenta lines represent the best linear model fit with $r\in[20,70]$ Mpc/$h$.}
    \label{Figure5}
\end{figure}

Although this biasing scheme would work on lightcones by construction, for completeness we measure (see \autoref{sec:2lpt}), we measure the QSO 2PCF monopole and quadrupole\footnote{The 2PCF in light cones was computed using the public code \texttt{pycorr}: \url{https://github.com/cosmodesi/pycorr}.} directly on 10 \texttt{CoLoRe-2LPT} full sky $10$ Gpc/h light cones with $4096^{3}$ cells and $0 \leq z \leq 3.8$ meeting the final DESI mock requirements to complete the QSO clustering validation. We computed these for three redshift bins with approximately equal number of QSOs: $1.8-2.3$, $2.3-2.8$, $2.8-3.5$ and also for the whole light cone redshift range, resulting in an effective redshift of $z_{\rm{eff}}=2.33$. As an example, and to confirm that the shape of the 2PCF monopole and quadrupole do not present any ill behavior, we show the measurements at this $z_{\rm{eff}}$ in \autoref{Figure5}.

Finally, we fit the 2PCF monopole and quadrupole using linear theory as above, leaving the QSO bias as the only free parameter. This fit is represented with the dashed magenta lines in \autoref{Figure5} and is performed for distances within the range $20 \leq r\leq 70$ [Mpc/$h$]. For all cases, we obtain a QSO bias consistent with the observations within $1\%-5\%$ as shown in \autoref{Figure4}.

\subsection{Generating \texttt{CoLoRe-2LPT} lightcones}\label{sec:colorerun}
Having concluded our validation, we generate $400$ \texttt{CoLoRe-2LPT} full-sky light cones and skewers to be used in the subsequent sections of this work. These light cones are $\sim10$ Gpc/h with $\sim12$M QSOs covering the full DESI QSO redshift range $0 < z < 3.8$ and have a resolution of $4096^{3}$ cells of size $\sim 2.4$ Mpc/$h$. 

\section{\texttt{LyaCoLoRe-2LPT}: Creating Lyman-$\alpha$ transmission skewers}\label{sec:lyacolore}
Having generated a cosmological light cone realisation including QSOs and the underlying dark matter density and velocity fields, \texttt{CoLoRe-2LPT} can save this information along the LOS from each QSO to the central observer by performing a trilinear interpolation of both fields. These LOS fields contain all the cosmological information through their 3D large-scale clustering and are the seeds of our final Ly$\alpha$ transmission skewers. The procedure to transform these density and velocity skewers into Ly$\alpha$ transmission is carried out with the \texttt{python} module \texttt{LyaCoLoRe}\footnote{\url{https://github.com/igmhub/LyaCoLoRe/tree/master}.} \citep{LyaCoLoRe}. \texttt{LyaCoLoRe} has been well tested in the past for the case of log-normal realizations of \texttt{CoLoRe} with Gaussian skewers \citep{LyaCoLoRe, LauraCasas}. Here, we present and validate for the first time its pipeline adapted for the 2LPT skewers case: \texttt{LyaCoLoRe-2LPT}.
\subsection{From density to flux skewers}\label{sec:dens2flux}
The structure of \texttt{LyaCoLoRe} can be summarized in four main steps: adding small-scale power, converting density to optical depth, adding RSDs, and computing flux from optical depth. 

\subsubsection{Small scales}\label{sec:smallscale}
The main step of \texttt{LyaCoLoRe} is the addition of one dimensional small-scale power to \texttt{CoLoRe-2LPT}'s density skewers. This step is key to achieve the right 3D large-scale correlations together with the 1D small-scale clustering characteristic of the forest, down to scales of $\sim 100$ kpc/h \citep{Karacayli22}, which are one order of magnitude smaller than our mock resolution (see \autoref{sec:colorerun}). Although the fundamental cosmology (such as BAO or FS information) is already encoded in the \texttt{CoLoRe-2LPT}'s skewers, the 1D small-scales must still be realistic enough to match observational statistics like the 1D power spectrum ($P_{1D}$) up to $k_\parallel \lesssim 0.5$ $h$/Mpc.
This is essential because inaccuracies in small-scale modeling propagate into the covariance of subsequent 3D correlation measurements, affecting the error estimates even at large scales.

The procedure is as follows. First, we interpolate the \texttt{CoLoRe-2LPT} density and velocity skewers interpolated to a one-dimensional grid using the nearest grid point (NGP) scheme, with a pixel spacing of
$0.25$ Mpc/$h$ ($\delta_{2LPT}$). Second, we generate a set of independent small-scale Gaussian density contrast skewers ($\delta_{\varepsilon}$) in this grid according to the following one dimensional power spectrum:
\begin{equation}
P_{\rm 1D}(k) \propto \left[ 1 + \left( \frac{k}{k_1} \right)^n \right]^{-1}.
\end{equation}
This power spectrum mimics the shape of the observational $P_{\rm{1D}}$ presented in \citet{McDonald} and is normalized to have unit variance, so that we can control the amplitude of the extra power independently. The parameters $k_{1}$ and $n$ are free and we will tune them in \autoref{sec:tuning}.

We add this extra power to our original \texttt{CoLoRe-2LPT}'s skewers in the following manner:
\begin{equation}\label{EqDens}
\delta(z,\textbf{x}) + 1= (\delta_{2LPT}(z, \textbf{x})+1)(\delta_{\rm{LN,}\varepsilon}(z,\textbf{x})) + 1),
\end{equation}
where $\delta_{\rm{LN,}\varepsilon}$ represents the log-normal transformation of the small-scale field, i.e.:
\begin{equation}
\delta_{\rm{LN,}\varepsilon}(z,\textbf{x}) = \text{exp}\Big(D(z)\sigma_{\varepsilon}(z)\delta_{\varepsilon}(\textbf{x})-D^{2}(z)\frac{\sigma^{2}_{\varepsilon}(z)}{2}\Big),
\end{equation}
and $\sigma_{\varepsilon}(z)$ 
controls the amplitude of the small-scale fluctuations that will be fitted in \autoref{sec:tuning} 

\autoref{EqDens} ensures that our final $\delta$ arises from a physical density ($\rho(x) \geq 0$, $\forall x$) and couples the large-scale contrast with the small-scale one. Note that the shape of this expression arises directly from the fact that the \texttt{CoLoRe-2LPT} density skewers are 2LPT. For the log-normal case, where the skewers were Gaussian, this addition was simply a direct sum.

\subsubsection{Optical depth}\label{sec:opticaldepth}
Once our density contrast skewers have the correct clustering at both large and small scales, we need to generate the Ly$\alpha$ optical depths from the continuous \texttt{CoLoRe-2LPT} density fields. We rely on the Fluctuating Gunn-Peterson Approximation (FGPA) \citep{Bi}. By assuming photoionization equilibrium and a power-law intergalactic medium (IGM) temperature-density relation, the optical depth ($\tau$) is given by:
\begin{equation}
\tau(z,\textbf{x}) = \tau_{0}(z)(1+\delta(z,\textbf{x}))^{\alpha(z)},
\end{equation}
where $\tau_{0}$ captures the role of the gas temperature and photoionization in the approximation and $\alpha$ is given by the density-temperature relation of the IGM \citep{Bi, Weinberg, Croft, McQuinn}. These parameters are free and will be tuned in  \autoref{sec:tuning}.

\subsubsection{Redshift-space distortions}\label{sec:rsds}
The third step to generate transmission skewers is to apply redshift-space distortions (RSDs) to the optical depth. This accounts for the fact that observed redshifts are perturbed from the pure Hubble flow by the gas's gravitational and thermal peculiar velocities \citep{Kaiser, McQuinn}. The effect of RSDs is taken into account in \texttt{LyaCoLoRe-2LPT} by mapping real space optical depth into redshift space via the following expression\footnote{Note that this is the analytical expression of the mapping, the actual transformation is performed via a weight matrix like in \citet{LyaCoLoRe}.}:
\begin{equation}
\tau(s) = \int{\tau(x)\delta_{D}(s-x-a_{v}v)}\mathrm{d}x,
\end{equation}
where $s$ are redshift space coordinates, $x$ real space ones, $\delta_{D}$ denotes a Dirac delta, $v$ is the linear gravitational peculiar velocities of the skewers calculated by \texttt{CoLoRe-2LPT}\footnote{These velocities are rescaled in order to have the proper units (Mpc/$h$).}, and $a_{v}$ is a free parameter scaling the Ly$\alpha$ velocities. We ignore thermal broadening effects due to our resolution. The parameter $a_{v}$ will be tuned in \autoref{sec:tuning} to match the level of anisotropic clustering observed in Ly$\alpha$ data (quantified by $\beta_{\rm{Ly}\alpha}$). 

\subsubsection{Ly$\alpha$ Transmitted Flux}\label{sec:lyaflux}
Finally, we get the Ly$\alpha$ transmission via: 
\begin{equation}
F(s) = \text{exp}(\tau(s)),
\end{equation}
where $F$ is the transmitted flux in redshift space, which will be interpolated to a wavelength grid of 0.2 \AA{} resolution in the final mock -- roughly a fourth of a DESI pixel. Note that we chose this smaller value to avoid extra artifacts caused by the pixelization of the Ly$\alpha$ transmission files.

\subsection{Parameter tuning}\label{sec:tuning}
With the framework presented in the previous subsections, we are left with the following free parameters: $n$, $k_{1}$, $\sigma_{\varepsilon}(z)$, $\tau_{0}(z)$, $\alpha(z)$ and $a_{v}$. 
We tune these parameters to reproduce the latest DESI Ly$\alpha$ clustering statistics measurements, namely: the one dimensional power spectrum ($P_{\rm{1D}}$) \citep{Ravoux, Karacayli25}, the mean flux ($\overline{F}$) \citep{Turner}, the Ly$\alpha$ bias ($b_{\rm{Ly}\alpha}$) and the Ly$\alpha$ RSD parameter ($\beta_{\rm{Ly}\alpha}$) \citep{DESIII}. Remarkably, this is the first time mocks are fully fitted to DESI statistics, while previous Ly$\alpha$ mocks were tuned to BOSS/eBOSS. We aim to reproduce these quantities within $10\%$ of the observational measurements for $k_{\parallel} \lesssim 0.01$ s/km -- approximately $1$ $h$/Mpc at $z=1$ -- and redshift range of $2.2 \leq z \lesssim 3.5$.

We perform the calibration of these parameters in several steps. First, we follow the prescription of \cite{LyaCoLoRe} to tune $n$, $k_{1}$, $\sigma_{\varepsilon}(z)$ and $\tau_{0}(z)$, keeping $\alpha(z)=1.65$ fixed across redshifts \citep{SeljakBias} during the complete calibration process and $a_{v}=1$ until the last step of the process. 

Second, given that the procedure of \cite{LyaCoLoRe} was not designed for 2LPT skewers, specifically for the computation of $b_{\rm{Ly}\alpha}$, we perform a second calibration using as initial values the ones obtained in the first step. In this second calibration, we keep $n$ and $k_{1}$ fixed, as they are mainly constrained by the size of the \texttt{CoLoRe-2LPT} cell, and vary $\tau_{0}(z)$ and $\sigma_{\varepsilon}(z)$ at discrete redshift bins\footnote{We applied these variations to the mock smoothly with redshift via a cubic spline interpolation assuming that the redshift dependence asymptotically converges to one assumed in \cite{LyaCoLoRe}.} ($z=2.2, 2.4, 2.8$ and $3.2$) covering the range of DESI measurements. We create a $8\times8$ regular grid of \texttt{CoLoRe-2LPT} mocks varying these parameters up to $20\%$ of the initial values, measure $P_{\rm{1D}}$, $\overline{F}$ and $b_{\rm{Ly}\alpha}$ (see \autoref{sec:lya_val}) and study which variation of the parameters best reproduce the DESI measurements.

Finally, once we have calibrated all other parameters, we manually vary $a_{v}$ between $[1.0,1.3]$ with a resolution of $0.05$ until we find a good agreement with the observational $\beta_{\rm{Ly}\alpha}$. Note that we expect $a_{v}$ to differ from unity to reproduce the anisotropies observed in the forest. This is not a sign that the \texttt{CoLoRe-2LPT} velocities are wrong, but rather a consequence of the approximations involved in obtaining $F$ which directly affect the large-scale forest biases. The final values of all these tuning parameters are public in the \texttt{LyaCoLoRe} repository\footnote{\url{https://github.com/igmhub/LyaCoLoRe/blob/master/input_files/tuning_files/tuning_data_2lpt_v2.8.fits}.}.

\section{Validation of the Lyman-$\alpha$ forest clustering statistics in \textit{raw} mocks}\label{sec:lya_val}
This section validates the procedures outlined in \autoref{sec:lyacolore} by confirming that we recover the DESI Lyman-$\alpha$ clustering statistics in \textit{raw} mocks, i.e., mocks that do not include noise or contaminants. Specifically, we present $P_{\rm{1D}}$ and $\overline{F}$ in \autoref{sec:1d_val}, and $b_{\rm{Ly}\alpha}$ and $\beta_{\rm{Ly}\alpha}$ in \autoref{sec:lya_auto}. We treat the biases separately because extracting them in the 2LPT framework requires computing and fitting the Ly$\alpha$ auto-correlation and the Ly$\alpha$ $\times$ QSO cross-correlation. Furthermore, analyzing these correlations in greater detail demonstrates the suitability of the mocks for BAO and FS analyses, while emphasizing the non-linear improvements introduced by the 2LPT method with respect to previous approaches.

\subsection{The one dimensional power spectrum and mean flux}\label{sec:1d_val}
We compute the $P_{\rm{1D}}$ and $\overline{F}$ using the Package for IGM Cosmological-Correlations Analyses (\texttt{picca}\footnote{\url{https://github.com/igmhub/picca/}.}) for $500,000$ independent \texttt{LyaCoLoRe-2LPT} skewers with the final parameter configuration of \autoref{sec:tuning}.

\begin{figure}[h]
    \centering
    \includegraphics[width=\linewidth]{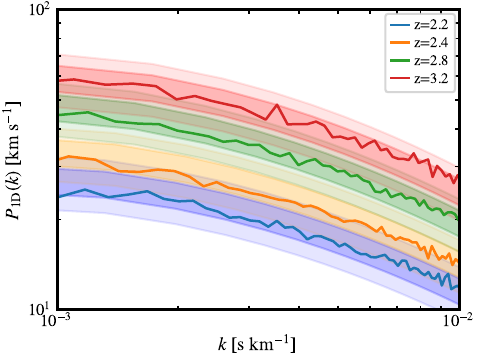}
    \caption{\texttt{LyaCoLoRe-2LPT} $P_{\rm{1D}}$ for $500,000$ skewers at the main redshift bins covering the DESI Ly$\alpha$ range ($z=2.0,2.5,2.8,3.2$) and for the relevant $k$ modes for BAO analysis ($k\in[10^{-3},10^{-2}]$ km/s). We compare the \texttt{LyaCoLoRe-2LPT} result against the DESI DR1 measurement \citep{Ravoux, Karacayli25} whose $10\%$ and $20\%$ are represented as strong and light shadows.}
    \label{Figure6}
\end{figure}

In \autoref{Figure6} we compare the measured $P_{\rm{1D}}$ in the mocks -- as solid lines -- against the best fit of DESI DR1 $P_{\rm{1D}}$ \citep{Ravoux, Karacayli25} -- as shades -- for the DESI Ly$\alpha$ redshift range and the relevant $k$ modes for our validation (see previous sections). The darker shades represent the $10\%$ of the DESI measurement and the light shades the $20\%$.

\begin{figure}[h]
    \centering
    \includegraphics[width=\linewidth]{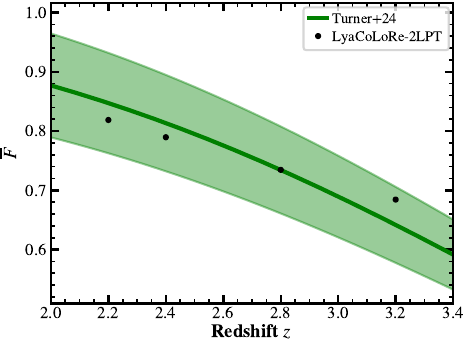}
    \caption{Comparison of \texttt{LyaCoLoRe-2LPT} $\overline{F}$ for $500,000$ skewers at the main redshift bins covering the DESI Ly$\alpha$ range ($z=2.0,2.5,2.8,3.2$) against the DESI DR1 measurement \citep{Turner} whose $10\%$ is represented as a shadow.}
    \label{Figure7}
\end{figure}

We show the results equivalently for the mean flux in \autoref{Figure7} against the DESI results from \citet{Turner} and its $10\%$ error. In both cases, we observe that the performed tuning is efficient, with statistics lying within DESI's $10\%$ for almost all cases.

\subsection{The Lyman-$\alpha$ auto-correlation and cross-correlation with QSOs}\label{sec:lya_auto}
We compute and fit the 3D Ly$\alpha$ auto-correlation and Ly$\alpha$ $\times$ QSO cross-correlation following the procedures of the latest Ly$\alpha$ analysis of DESI \citep{DESIII, CuceuFS2} in 10 full sky mocks (equivalent to $\sim 70$ DESI realizations) of the 400 \textit{raw} realizations described in the previous section. That is, we will compute these correlations for 3,072 \texttt{HEALPix} \citep{HealpixRef} pixels in the sky that will be treated as 30,720 subsamples for our 10 realizations. The covariance matrix of the correlations will be computed directly by a smoothed subsampling of these set of independent realizations as done for the observations\footnote{All the correlations and covariances are computed using \texttt{picca}}.

Both the 3D Ly$\alpha$ auto-correlation and Ly$\alpha$ $\times$ QSO cross-correlation will be computed as a weighted average following:
\begin{align}\label{eq:average}
    \xi^{\rm auto}_{A} &= \frac{\sum_{(i,j)\in A}w_{i}w_{j}\delta^{i}_{F}\delta^{j}_{F}}{\sum_{(i,j)\in A}w_{i}w_{j}},
    &
    \xi^{\rm cross}_{A} &= \frac{\sum_{(i,j)\in A}w_{i}w_{j}\delta^{i}_{F}}{\sum_{(i,j)\in A}w_{i}w_{j}},
\end{align}

where $\delta_F$ represents the flux contrast:
\begin{equation}\label{eq:contrast}
    \delta_{F}(\lambda) = \frac{F(\lambda)}{\overline{F}(\lambda)}-1,
\end{equation}
and $w$ are weights defined as:
\begin{equation}
w_{i} = \Big(\frac{1+z_{i}}{1+z_{\rm{eff}}}\Big)^{\gamma-1},
\label{eq:weights}
\end{equation}
where $z_{i}$ describes the redshift of the pixel, $z_{\rm{eff}}$ is the effective redshift bin -- for our realizations $z_{\rm{eff}} = 2.33$ -- and $\gamma$ is a parameter characterizing the redshift evolution of the tracer -- for the Ly$\alpha$ case, $\gamma_{\text{Ly}\alpha}=2.9$, while for the QSO case, $\gamma_{\rm{QSO}} = 1.44$. Note that for the cross-correlation the $\delta^j_{F}$ term in the auto-correlation expression is dropped due to the discreteness of QSOs \citep{AndreuDLAs}.

In both cases, we restrict to the flux contrast in pixels with rest-frame wavelength inside the Ly$\alpha$ forest: $\lambda \in [1040, 1205]$ \AA{} and resample our skewers to pixels of width $3 \times 10^{-4}$ log(\AA) in log-wavelength for computational efficiency \citep{LyaCoLoRe}.

The auto-correlation is computed for every pair of pixels $(i,j)$ of parallel and perpendicular separation $(r_{\parallel}, r_{\perp}) \in [0,200]$ Mpc/$h$ in both directions, in bins of $4$ Mpc/$h$, denoted with $A$ in \autoref{eq:average}. For the cross-correlation, $r_{\parallel}$ ranges from $-200$ to $200$ Mpc/$h$ -- with  $r_{\parallel}>0$ for Ly$\alpha$ pixels with $z < z_{\rm{QSO}}$ for the considered QSO.

We fit the computed auto- and cross-correlation relating the Ly$\alpha$ and QSO power spectra $\hat{P}_{ij}(\textbf{k})$ with the underlying dark matter power spectrum $P(\textbf{k})$ as:

\begin{equation}\label{eq:pkmodel}\hat{P}_{ij}(\textbf{k}) = b_{i}b_{j}(1+\beta_{i}\mu_{k}^{2})(1+\beta_{j}\mu_{k}^{2})G(\textbf{k})P(\textbf{k}),
\end{equation}
where $(i,j) \in \{(\rm{Ly}\alpha,\rm{Ly}\alpha),(\rm{Ly}\alpha,\rm{QSO})\}$, $\textbf{k} \equiv(k_{\parallel},k_{\perp}) \equiv (|\textbf{k}|, \mu_{k})$, 
$\mu_{k} = |\textbf{k}|/k_{\parallel}$ is the cosine of the angle to the LOS, $b_{i}$ is the bias of the tracer $i$ and $\beta_{i}$ characterizes the anisotropic clustering of the tracer $i$: $\beta_{i}:= fb_{\eta,i}/b_{i}$, where $f$ is the cosmological growth rate and $b_{\eta,i}$ is the velocity gradient bias of the tracer $i$. For QSOs this gradient velocity bias is unity since their number is conserved under RSD transformation due to discreteness. The term $G(\textbf{k})$ is a window function to take into account the grid binning when computing the correlations. In the modelling presented in this section we focus on Kaiser models, avoiding any terms related to contaminants or distortion by continuum since we are working on \textit{raw} mocks. These extra terms in the modelling will be studied in \autoref{sec:val_cont}.

We consider the following decomposition of the fiducial power spectrum:
\begin{equation}
\begin{aligned}
P(\textbf{k}) &=  P_{\rm peak}(\textbf{k})
\exp\!\frac{\left(-k^{2}_\parallel \Sigma^{2}_\parallel
            -k^{2}_\perp \Sigma^{2}_\perp\right)}{2}
+ P_{\rm sm}(\textbf{k})
\exp\!\frac{\left(-k^{2}_\parallel \sigma^{2}_\parallel
            -k^{2}_\perp \sigma^{2}_\perp\right)}{2} \\
&= P_{\rm peak}(\textbf{k}) F_{\rm broad}(\textbf{k})
+ P_{\rm sm}(\textbf{k}) F_{\rm sm}(\textbf{k}),
\end{aligned}
\end{equation}
this is, a smooth and a peak component defining BAO. The term $F_{\rm{broad}}$ accounts for the non-linear broadening of the BAO peak and the term $F_{\rm{sm}}$ for non-linearities that cannot be captured directly by the model due to the resolution of the \texttt{CoLoRe-2LPT} cells. Hence, these $\sigma_{\perp, \parallel}$ are expected to be of the order of this size, $\sim 2.4$ Mpc/$h$. 

Finally, this Fourier space relation is transformed to configuration space where the actual fit is being carried out as:
\begin{equation}\label{eq:ximodel}
    \xi_{ij}(r_\parallel,r_\perp) = \xi_{\rm{peak}}(q^{\rm{peak}}_{\parallel}r_\parallel,q^{\rm{peak}}_{\perp}r_\perp) +  \xi_{\rm{sm}}(q^{\rm{sm}}_{\parallel}r_\parallel,q^{\rm{sm}}_{\perp}r_\perp),
\end{equation}
where $q_{\parallel}$ and $q_{\perp}$ are the coefficients capturing cosmology via the rescaling of coordinates\footnote{Note that for all parameters related to these $q$ rescaling parameters we will use the superindeces peak and sm to differentiate which component of the correlation was used to fit them, in contrast to no super-index denoting that the full-shape of the correlation was used, i.e. we force peak=smooth.}. 

\begin{figure*}[t]
    \centering
    \includegraphics[width=\textwidth]{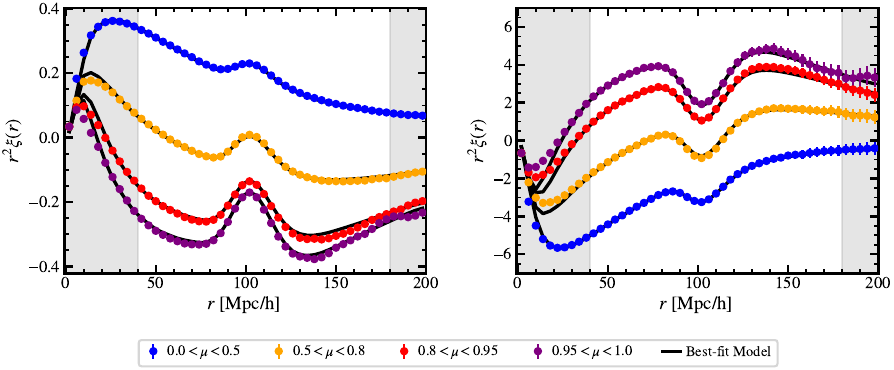}
    
    \caption{\textit{Left panel}: \textit{Raw} Ly$\alpha$ auto-correlation. 
    \textit{Right panel}: \textit{Raw} Ly$\alpha\times$QSO cross-correlation. In both panels, the correlations, presented as dots, represent the stack of 10 \textit{raw} \texttt{LyaCoLoRe-2LPT} full-sky mocks. The lighter dots are the individual mocks correlations and the solid lines represent the best FS linear fit for $40 < r$ [Mpc/$h$] $< 180$. The best BAO linear fit is indistinguishable from the FS (see \autoref{tab:fits}). The grey bands indicate the scales which are not included in the model fit}
    \label{fig:stack_raw}
\end{figure*}

For a BAO only analysis, $q^{\rm{sm}}_{\parallel}=q^{\rm{sm}}_{\perp}=1$ and the goal is to fit $(q^{\rm{peak}}_{\parallel},q^{\rm{peak}}_{\perp})$ which we will call $(\alpha_{\parallel},\alpha_{\perp})$.

For a FS analysis, we focus at the same time on BAO and the Alcock-Paczynski effect (AP) 
\citep{CuceuFS1, CuceuFS2}. In this case, we aim to fit $(q_{\parallel}$, $q_{\perp})$ with the parameterization:
\begin{equation}\label{eq:fsparam}
\phi(z) \equiv \frac{q_\perp(z)}{q_\parallel(z)}, 
\qquad
\alpha(z) \equiv \sqrt{q_\perp(z)\,q_\parallel(z)},
\end{equation}

We perform BAO and FS fits on the \textit{raw} mocks and compare the mock biases and cosmology results with the latest DESI DR2 BAO results \citep{DESIII}.  

For these fits\footnote{All the fits performed in the paper are carried out using the public code \texttt{vega}: \url{https://github.com/andreicuceu/vega}.}, we use both the auto- and cross-correlation combined\footnote{We also computed and fitted the auto- and cross-correlation individually and find consistent results.}, using the previously described BAO and FS parameterizations while leaving free the Ly$\alpha$ biases $(b_{\rm{Ly}\alpha},\beta_{\rm{Ly}\alpha})$, the QSO bias $b_{\rm{QSO}}$, the BAO broadening $(\Sigma_\perp, \Sigma_\parallel)$ and smoothing $(\sigma_\perp, \sigma_\parallel)$ parameters. The fits are carried out in the range $40$ $\leq r$ [Mpc/$h$] $\leq 180$. We choose this $r_{min}$ to be conservative, since we do not expect the presented linear model to capture any non-linearity present in the smaller scales of the new 2LPT mocks.

We show the auto- and cross-correlation measurements for the stack of 10 \texttt{LyaCoLoRe-2LPT} \textit{raw} full-sky realizations as dots for different $\mu$ wedges and the best FS fit combined model as solid lines in \autoref{fig:stack_raw}. We find an indistinguishable best-fit for BAO. As showed, the model fits the data points almost perfectly\footnote{Note that the $\chi^{2}$ values are not representative of data at this stage, because we are working on \textit{raw} mocks that do not have instrumental noise.} for the $r$ range included in the fit, only failing in smaller separations which are not included. We present the FS and BAO best-fit parameters in \autoref{tab:fits}.

\begin{table}[!htbp]
\centering
\small
\caption{Parameters from the best Full Shape and BAO fits to the combination of the Ly$\alpha$ auto-correlation and Ly$\alpha$ $\times$ QSO cross-correlation from the stack of $10$ \textit{raw} realizations of \texttt{LyaCoLoRe-2LPT}. We present the DESI DR2 BAO
values \cite{DESIII} to which the Ly$\alpha$ bias and $\beta$ parameter are tuned following the procedure of \autoref{sec:tuning}.}
\label{tab:fits}

\resizebox{\columnwidth}{!}{
\begin{tabular}{lccc}
\hline\hline
 & Full Shape & BAO & DESI DR2 BAO \\
\hline
$b_{\mathrm{Ly}\alpha}$     & $-0.1444 \pm 0.0006$   & $-0.1439 \pm 0.0005$ & $-0.1352 \pm 0.0073$ \\
$\beta_{\mathrm{Ly}\alpha}$ & $1.346 \pm 0.008$ & $1.347 \pm 0.008$ & $1.445 \pm 0.064$ \\
$b_{\mathrm{QSO}}$          & $3.405 \pm 0.014$ & $3.394 \pm 0.007$ & $3.545 \pm 0.054$ \\
$\Sigma_{\parallel}$        & $6.46 \pm 0.10$ & $6.45 \pm 0.09$ & -- \\
$\Sigma_{\perp}$            & $3.40 \pm 0.18$ & $3.29 \pm 0.17$ & -- \\
$\sigma_{\parallel}$        & $3.11 \pm 0.12$ & $3.05 \pm 0.10$ & -- \\
$\sigma_{\perp}$            & $3.02 \pm 0.16$ & $2.98 \pm 0.13$ & -- \\

$\alpha_{\parallel}$        & -- & $1.003 \pm 0.001$ & --  \\
$\alpha_{\perp}$ & -- & $1.000 \pm 0.001$ & -- \\

$\phi$ & $0.999 \pm 0.001$ & -- & -- \\
$\alpha_{\rm peak}$ & $1.001 \pm 0.001$ & -- & -- \\
$\alpha_{\rm smooth}$ & $1.002 \pm 0.002$ & -- & -- \\
\hline
\hline
\end{tabular}
}
\end{table}

As shown in \autoref{tab:fits}, the BAO and FS cosmological parameters deviate from unity by $0.003$ at most. Given that the statistical uncertainties for DESI DR2 are an order of magnitude larger, this minor systematic shift is negligible, rendering the mocks effectively unbiased. Moreover, demonstrating the complete validity of the tuning process (\autoref{sec:tuning}), the Ly$\alpha$ bias lies within $\sim0.4\sigma$ of the DESI DR2 BAO value. The $\beta_{\rm{Ly}\alpha}$ parameter is also consistent within $\sim1\sigma$, demonstrating the good choice of the velocity scaling parameter $a_{v}$. On its side, the QSO bias, though lower than the DESI DR2 Ly$\alpha$ BAO value, is consistent with the input value used in \autoref{sec:qso_clustering} and, consequently, the DESI DR1 measurement \citep{ChaussidonQSO}. Finally, the smoothing parameters are consistent with the cell size of \texttt{CoLoRe-2LPT} and the BAO broadening parameters are consistent with the expected theoretical values -- $\Sigma_{\parallel}\sim6.4$$h$/Mpc and $\Sigma_{\perp}\sim3.26$ $h$/Mpc \citep{Eisenstein2007, Kirkby2013}. This broadening arises from large-scale bulk flows, which smear the acoustic feature with an rms given by the Zel'dovich displacement of the linear density field, enhanced along the line of sight by RSD. Since 2LPT captures these displacements, our mocks naturally reproduce the anisotropic damping measured in Ly$\alpha$ simulations \citep{Boryana, BoryanaShift, Francesco2}, in contrast with the ad-hoc broadening of previous lognormal or quasi-linear mocks \citep{LyaCoLoRe, LauraCasas}.

In this line, we also tried to fit the model up to smaller scales ($r_{min}<40$ Mpc/$h$). We find that since the used model is mainly based on linear theory, the fit starts to fail at 30 Mpc/$h$ due to the non-linearities of \texttt{LyaCoLoRe-2LPT}, while the previous \texttt{CoLoRe-QL} \citep{LauraCasas} could be fitted  down to scales of $10$ Mpc/$h$. This demonstrates the greater realism of these mocks, which include better non-linearities. Furthermore, if a non-linear model correction \citep{Arinyo2015} is included, the fit is improved for all $r_{min}$, while no improvement was observed for the previous log-normal mocks. These tests will be presented with more detail in the companion papers \citep{DESI:2026DR2IV, MHVal} for the validation of the DESI DR2 FS analysis. 

We can conclude that the mocks are sufficient for DESI BAO and FS analysis since they reproduce the observational clustering statistics of the Ly$\alpha$ forest and QSOs and include better non-linearities than previous mocks, fulfilling their design aim.

\section{Generating \textit{contaminated} DESI-like Ly$\alpha$ mocks}\label{sec:gen_desi}
While in the previous sections we have described and validated \textit{raw} mocks, we now aim to generate mocks that closely resemble the DESI DR2 data -- including contaminants, continuum fitting and its associated distortion -- for the proper validation of the observational analysis.

\subsection{Adding Astrophysical Contaminants}\label{sec:contaminants}
The first step towards mocks that resemble data is including any physical contaminant that can produce systematics and bias the desired analysis. For Ly$\alpha$ forest analyses, the two main contaminants are High Column Density Systems (HCDs) and metals. 

\subsubsection{High Column Density Systems (HCDs)}\label{sec:hcd}

While in previous sections we have primarily modeled the Ly$\alpha$ forest as diffuse IGM absorption, the LOS can intersect denser and more clustered environments -- e.g. the circumgalactic medium -- originating High-Column Density systems (HCDs). These include, depending on their neutral hydrogen column density ($N_{\rm{HI}}$): Damped Ly$\alpha$ systems (DLAs, $N_{\rm HI} > 2 \times 10^{20}$ cm$^{-2}$) and Lyman Limit Systems (LLS, $10^{17.2} < N_{\rm HI} < 2 \times 10^{20}$ cm$^{-2}$) \citep{Wolfe, McQuinn}. Modeling HCDs is essential, as their broad absorption features bias the forest correlations \citep{AndreuHCDs, Bautista, RogersMocks}. Furthermore, they are of independent interest as tracers of the matter field \citep{AndreuDLAs, IgnasiBiasDLAs, IgnasiBiasSBLAs} and metal enrichment \citep{LluisMas, SeanSBLAs}.

We model HCDs by populating skewer cells\footnote{This procedure is performed on the $\textit{raw}$ \texttt{CoLoRe-2LPT} skewers without the small scale power of \autoref{sec:smallscale} since we want the HCDs to be correlated with the cosmological large-scale density field.} above a density threshold ($t_{\rm{HCD}}$) to match their clustering via linear bias ($b_{\rm{HCD}}$). For the 2LPT case the relation between $t_{\rm{HCD}}$ and $b_{\rm{HCD}}$ is not analytical, as presented in \autoref{sec:qso_clustering} for QSOs. Following the methods presented in that section, we derive this relation empirically. We generate 25 realizations of \texttt{CoLoRe-2LPT} snapshots at $z \in \{1.5, 2.0, 2.5, 3.0, 3.5$, $4.0$\} with $L_{\rm{box}} = 2$ Gpc/h and $n_{\rm{grid}}=1024$ matching the density field variance of the \autoref{sec:colorerun} \textit{raw} mocks. We populate these boxes using the biasing-scheme of \autoref{eq:biasing} leaving $t$ free and fixing $b=1$ for application at the skewer level. We run 120 realizations for $t \in (-0.6, 0.6)$ -- having previously discarded lower and higher values in coarser meshes -- and fit the HCD power spectrum to linear theory obtaining the best $t_{\rm{HCD}}(z)$ values that give a $b_{\rm{HCD}}\sim 2$ constant with redshift \citep{AndreuDLAs, IgnasiBiasDLAs, IgnasiBiasSBLAs}. The resulting best-fit thresholds redshift dependency is well described by the following empirical power-law relation:
\begin{equation}
t_{\rm{HCD}}(z) = a \, (1+z)^{b} + c,
\end{equation}
where the parameters are: $a = 5.331$, $b=-2.444$ and $c=-0.260$. We then compute the Ly$\alpha$ $\times$ HCD cross-correlation for 10 \textit{raw} full-sky realizations of \texttt{LyaCoLoRe-2LPT} to verify that the desired clustering is achieved, see \autoref{AppendixB} for details.

The second important point in HCD modeling is to match approximately their incidence rate ($dn/dz)$ in the Ly$\alpha$ skewers. Based on \texttt{pyigm}\footnote{\url{https://github.com/pyigm/pyigm}.} models for this $dn/dz$ as a function of column density \citep{ProchaskaDLA}, we populate the cells above the threshold with a Poisson distribution with mean:
\begin{equation}
\mu(z_{cell}) = \frac{dn/dz \: \Delta z_{\rm{cell}}}{p_{\nu}(z_{\rm{cell}})}.
\end{equation}
This is the mean number of HCDs per redshift resolution element ($\Delta z_{\rm{cell}}$) rescaled by the number of expected cells above the threshold at that redshift ($p_{\nu}$). Since the 2LPT density distribution is not analytical, we direcly compute this probability as the ratio of flagged cells with respect to total cells for each redshift resolution element. We calculate $p_{\nu}$ for one full-sky realization of \texttt{LyaCoLoRe-2LPT} and use it as an input for subsequent runs since its variability is negligible. We show the agreement between the measured $dn/dz$ of a \texttt{LyaCoLoRe-2LPT} full sky \textit{raw} mock and the \texttt{pyigm} model for different ranges of redshift and column density in \autoref{AppendixB}.

\subsubsection{Metal absorptions}\label{sec:metals}
The second major category of contaminants in Ly$\alpha$ forest analyses comprises metal line transitions. Because these lines possess different rest-frame frequencies than Ly$\alpha$, they masquerade as Ly$\alpha$ absorption originating from different redshifts. When computing line-of-sight distances under the assumption that all absorption is Ly$\alpha$, this frequency offset introduces spurious correlation bumps at specific comoving separations, making accurate metal modeling strictly necessary. 

We model these metal transitions using the same model as previous mocks assuming that the metals optical depth is linear to the Ly$\alpha$ one \citep{LyaCoLoRe, HiramQQ}. Note that differently from \citet{LyaCoLoRe}, we add the metal transitions after RSDs are included, using \texttt{quickquasars}\footnote{\url{https://github.com/desihub/desisim/blob/main/py/desisim/scripts/quickquasars.py}} \citep{HiramQQ}. Although it would be more realistic to add metals using a similar prospect to that of HCDs, that goes beyond the scope of this paper since we are solely interested in the shape and strength of the metal bumps in the correlations. This, together with a more detailed study of the HCD cross-correlation, will be presented in a separate paper (Ruiz-Herrera Bernal et al., in prep). The optical depth of each metal transition is then obtained as:
\begin{equation}
\tau_{\rm{metal}} = A_{\rm{metal}}\tau_{Ly\alpha},
\end{equation}
where $A_{\rm{metal}}$ is a free constant for each metal. These constants are tuned in order to obtain metal biases and correlation shapes consistent with those of DESI DR2 BAO \citep[][Fig. 3]{DESIII}. We model SiII(1190), SiII(1193), SiIII(1207) and SiII(1260) which are the principal transitions affecting the Ly$\alpha$ correlation. We find that the \texttt{quickquasars} $A_{\rm{metal}}$ values for the \texttt{CoLoRe-QL} mocks \citep{HiramQQ}  are suitable for the \texttt{LyaCoLoRe-2LPT} mocks. We show the agreement of the metal biases with the DESI DR2 BAO values and the similarity of the correlation functions compared to data in \autoref{sec:val_cont}.

\subsection{Adding QSO continua and noise}\label{sec:QQ}
After generating the \textit{raw} mocks and defining the modeling of the main Ly$\alpha$ astrophysical contaminants, we can produce DESI-like \textit{contaminated} mocks. 

We produce these \textit{contaminated} mocks in three steps. First, we use the package \texttt{desisim}\footnote{\url{https://github.com/desihub/desisim}} to match the DESI DR2 footprint, number of QSOs and exposures, and randomly assign a magnitude to each QSO. Then, we add the previously described contaminants (HCDs and metals) together with Broad Absorption Lines (BALs) using the \texttt{python} script \texttt{quickquasars} \citep{HiramQQ}. Unlike for DLAs and metals, we do not describe the modeling of BALs in detail, since the prescription is identical to that used in past mocks \citep[see e.g.][]{HiramQQ, LauraCasas}. This script can also rescale the transmitted flux by a random unabsorbed quasar spectra (hereafter, \textit{continua}, $C(\lambda)$) generated using the \texttt{SIMQSO}\footnote{\url{https://github.com/imcgreer/simqso}} package \citep{McGreer2021}. Finally, we add DESI-like noise to the spectra using the package \texttt{specsim}\footnote{\url{https://github.com/desihub/specsim}}\citep{Kirkby2016}. We generate 400 realization of \texttt{LyaCoLoRe-2LPT} \textit{contaminated} mocks for this work and the validation of the DESI DR2 FS analysis presented in the companion papers \citep{DESI:2026DR2IV, MHVal}.

\section{Consistency of \textit{contaminated} mocks with DESI DR2 BAO measurements}\label{sec:val_cont}
In this section, we validate the \textit{contaminated} DESI-like mocks by fitting them with the full DESI DR2 BAO modeling and comparing the results directly with the observational measurement. The equivalent fit for FS is presented in the companion paper \citep{DESI:2026DR2IV}.

We compute the 3D Ly$\alpha$ auto-correlation and Ly$\alpha$ $\times$ QSO cross-correlation for these 400 realizations with the approach described in \cite{DESIII}. This procedure is exactly the same as the one described in \autoref{sec:lya_auto} with the particularity that we have to take into account the QSO continua. This modifies \autoref{eq:contrast} to:
\begin{equation}
    \delta_{F}(\lambda) = \frac{F(\lambda)}{\overline{F}C(\lambda)}-1,
\end{equation}
being a \textit{continuum fitting}, i.e., estimating $\overline{F}C(\lambda)$ for each QSO, required to compute $\delta_{F}$\footnote{Note that, in order to mimick the observational procedure, DLAs and BALs are masked prior to compute $\delta_{F}$. See \citet{LauraCasas, DESIII} for more details.}. We perform continuum fitting following the procedures described in \citet{eBOSSBAO} and \citet{CesarEDR}. This fitting suppresses long wavelength modes distorting the correlation functions. We correct for this effect by removing the first and mean moments from each spectra using the modified version of \autoref{eq:weights} presented in \citet{Busca25}.

\begin{figure*}[!t]
    \centering
    \includegraphics[width=\textwidth]{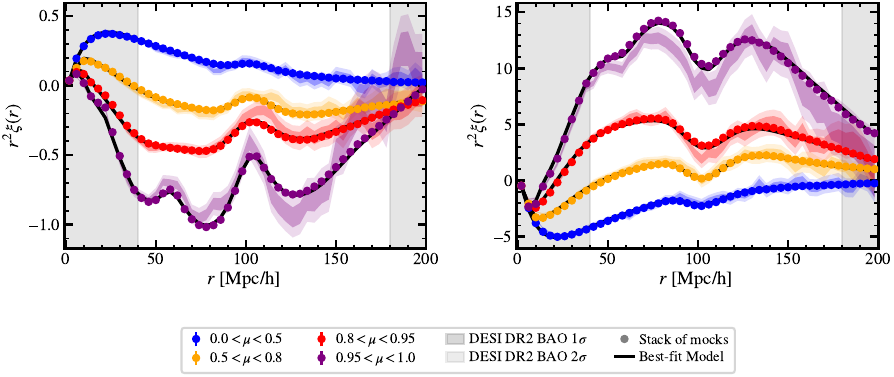}
    
    \caption{ \textit{Left panel}: \textit{Contaminated} Ly$\alpha$ auto-correlation. 
    \textit{Right panel}: \textit{Contaminated} Ly$\alpha\times$QSO cross-correlation. In both panels, the correlations presented as dots represent the stack of 400 \textit{contaminated} \texttt{CoLoRe-2LPT} DESI-like mocks. The solid lines represent the best BAO linear fit for $40 < r$ [Mpc/$h$] $< 180$ and the shadows represent the $1\sigma$ and $2\sigma$ errors of DESI DR2 BAO correlations. The grey bands indicate the scales which are not included in the model fit.}
    \label{colore_cont_corr}
\end{figure*}

We present the measured correlations for the stack of the 400 \texttt{LyaCoLoRe-2LPT} \textit{contaminated} realizations -- as points -- together with the DESI DR2 correlations -- as bands -- for different $\mu$ wedges in \autoref{colore_cont_corr}. The mock correlations strongly agree with the observations for all $r$, including the smaller scales. As described in \autoref{sec:metals}, we can see the spurious correlation signal from metal transitions as bumps at scales different from those of BAO in the $\mu\sim1$ wedge of the Ly$\alpha$ auto-correlation. The good agreement of these bumps between the mock and the data correlation qualitatively validates the metal modeling described in \autoref{sec:metals}. The cross-correlation also resembles the observational measurement, although, in this case, we find a stronger disagreement for the wedge closer to the LOS for $r\sim60-70$ Mpc/$h$. We attribute these discrepancies to the approximate modeling of HCDs (see \autoref{AppendixB}) and assuming Gaussian redshift errors which strongly affect the correlations near the LOS. Any other minor difference between the mock and data correlations is likely due to the non-perfect modeling of the tracers biases in strongly non-linear scales (as seen in previous mocks, e.g. \citet[][]{LyaCoLoRe,LauraCasas}). 

We perform the final validation of the \textit{contaminated} mocks by fitting them with the full\footnote{The only features we don't include in the fit are effects which are not modeled in the mocks like instrumental systematics or QSO radiation.} DESI DR2 BAO model. This model follows the same prescription as the one presented in \autoref{sec:lya_auto} based on \autoref{eq:pkmodel}. Here, we account for the impact of the contaminants and continuum fitting in the model. 

The metal contamination is included in the model taking into account the Ly$\alpha$-, QSO- and metal-metal correlations:
\begin{equation}
    \hat{\xi}_{ij} = \xi_{ij} + \sum_{m}\xi_{im}+\delta_{ij}\sum_{m,m'}\xi_{mm'},
\end{equation}
where $(i,j) \in \{(\rm{Ly}\alpha,\rm{Ly}\alpha),(\rm{Ly}\alpha,\rm{QSO})\}$, $m$ represents the different metal species of \autoref{sec:metals} and $\delta_{ij}$ is a Kronecker delta. Each metal species will be characterized by a $b_{m}$ and $\beta_{m}$ according to \autoref{eq:pkmodel}. We leave $\beta_{m}$ free and common for all metal species, as this parameter is not constrained due to the addition of metal absorption after RSDs. The line misidentification, coming from the assumption that all absorptions have Ly$\alpha$ rest-frame wavelength, is accounted for in the correlations including metals with the \citet{MetalMatrix} metal matrix model modified as in \citet{DESIII}. 

Regarding HCDs, we model the effect of non-masked systems using the Rogers model \citep{RogersMocks, eBOSSBAO}, which relates the effective measured $\hat{b}_{\rm{Ly}\alpha}$ and $\hat{\beta}_{\rm{Ly}\alpha}$ with the real Ly$\alpha$ and HCD biases introducing a scale-dependency via:
\begin{equation}
\begin{cases}
\hat{b}_{\rm{Ly}\alpha} = {b}_{\rm{Ly}\alpha} + b_{\rm{HCD}}\rm{exp}(-L_{\rm{HCD}}k_{\parallel}), \\
\hat{b}_{\rm{Ly}\alpha}\hat{\beta}_{\rm{Ly}\alpha} = {b}_{\rm{Ly}\alpha}{\beta}_{\rm{Ly}\alpha} + b_{\rm{HCD}}\beta_{\rm{HCD}}\rm{exp}(-L_{\rm{HCD}}k_{\parallel}),
\end{cases}
\end{equation}
where $L_{\rm{HCD}}$ depends on the column density of the unmasked systems.

Finally, we include the continuum fitting distortion by multiplying the measured correlation by the \textit{distorsion matrix} based on \citet{Busca25}. 

We fit the previously computed correlations following \autoref{eq:ximodel} for BAO including the contaminant modeling. We perform a combined BAO fit of the Ly$\alpha$ auto-correlation and the Ly$\alpha$ $\times$ QSO cross-correlation leaving free the parameters described in \autoref{sec:lya_auto} (\autoref{tab:fits}) and the metal biases $b_{\rm{m}}$ with $\rm{m}\in\{$SiII(1190), SiII(1193), SiIII(1206) and SiII(1260)$\}$. For simplicity, given the limitations of our HCD modeling, we perform a separate fit for HCD absorption (see Appendix \ref{AppendixB}) and fix the HCD parameters to those best fit values. This avoids degeneracies and potential problems in the model that can interfere with the validation of the mock methodology itself. This BAO fit is computed for $40 < r$ [Mpc/$h$] $<180$, consistent with the \textit{raw} mocks measurement (\autoref{sec:lya_auto}).

We present the best combined BAO fit combined as solid lines in \autoref{colore_cont_corr} and the best-fit parameters in \autoref{tab:bao_fits}. The model and the mock data points present a very good agreement for the separations included in the fit. The common parameters between the \textit{raw} (see \autoref{tab:fits}) and \textit{contaminated} case are consistent with each other, leading to equivalent conclusions regarding biases and non-linearities as those of \autoref{sec:lya_auto}. However, as found in the companion paper \citep{DESI:2026DR2IV} we find biased cosmology results for BAO at around $0.3\%$. This bias appears to be associated with the model itself and will be studied in more detail in the aforementioned companion paper and in the context of DESI DR3 BAO analysis. For the metal biases we find an agreement within $\sim20\%$ between \texttt{LyaCoLoRe-2LPT} and the DESI DR2 BAO parameters, supporting the validity of the metal modeling of \autoref{sec:metals}.

\begin{table}[!htbp]
\centering
\caption{Parameters from the best BAO fit to the combination of the Ly$\alpha$ auto-correlation and Ly$\alpha \times$QSO cross-correlation from the stack of 400 \textit{contaminated} realizations of \texttt{LyaCoLoRe-2LPT}. We present the DESI DR2 BAO values \citep{DESIII} to which the Ly$\alpha$ bias and $\beta$ parameter, and the HCD and metal biases are tuned following the procedures of \autoref{sec:tuning} and \autoref{sec:contaminants} respectively. $^{*}$Note that the HCD parameters are fixed to the best fit values encountered in Appendix \ref{AppendixB}.}
\label{tab:bao_fits}
\begin{tabular}{lcc}
\hline\hline
 & Mock BAO & DESI DR2 BAO \\
\hline
$b_{\mathrm{Ly}\alpha}$     & $-0.1434 \pm 0.0001$ & $-0.1352 \pm 0.0073$ \\
$\beta_{\mathrm{Ly}\alpha}$ & $1.3283 \pm 0.0012$ & $1.445 \pm 0.064$  \\
$b_{\mathrm{QSO}}$          & $3.2693 \pm 0.0014$ & $3.545 \pm 0.054$ \\
$b_{\mathrm{HCD}}$          & $-0.035^{*}$ & $-0.0206 \pm 0.0090$ \\
$\beta_{\mathrm{HCD}}$      & $0.485^{*}$ & $0.508 \pm 0.089$ \\
$L_{\mathrm{HCD}}$          & $5.512^{*}$ & $5.30 \pm 0.93$ \\
$10^{3}b_{\mathrm{SiII}\,(1190)}$   & $-3.312 \pm 0.024$ & $-3.70 \pm 0.39$\\
$10^{3}b_{\mathrm{SiII}\,(1193)}$   & $-1.679 \pm 0.016$ & $-3.18 \pm 0.38$ \\
$10^{3}b_{\mathrm{SiIII}\,(1206)}$  & $-9.84 \pm 0.08$ & $-7.3 \pm 1.5$ \\
$10^{3}b_{\mathrm{SiII}\,(1260)}$   & $-3.359 \pm 0.023$ & $-3.67 \pm 0.40 $  \\
$\beta_{m}$   & $1.005 \pm 0.017$ & $0.5 \text{ (fixed)} $  \\
\hline
\hline
$\alpha_{\parallel}$        & $1.0042 \pm 0.0005$ & $1.002 \pm 0.011$ \\
$\alpha_{\perp}$            & $0.9985 \pm 0.0006$ & $0.995 \pm 0.013$ \\
$\Sigma_{\parallel}$        & $6.09 \pm 0.07$  & -- \\
$\Sigma_{\perp}$            & $3.10 \pm 0.13$ & -- \\
\hline
\hline
\end{tabular}
\end{table}

The high level of agreement between the \texttt{LyaCoLoRe-2LPT} correlations and the DESI DR2 BAO, together with the results of \autoref{sec:lya_auto}, supports the robustness of these mocks for DESI cosmological analyses.

\section{Summary \& Conclusions}\label{sec:conclusions}

We have created a new generation of Lyman-$\alpha$ (Ly$\alpha$) forest mock catalogues, based on \texttt{CoLoRe-2LPT}, designed to validate DESI Baryonic Acoustic Oscillations (BAO) and Full-Shape (FS) cosmological analyses. These new mocks replace the log-normal density evolution used in previous \texttt{CoLoRe} implementations with a second‑order Lagrangian perturbation theory (2LPT) description. This approach improves the mildly non-linear scales while preserving the computational efficiency required to generate large ensembles, in contrast to N-body or hydrodynamical simulations.

First, we validated the 2LPT implementation in \texttt{CoLoRe-2LPT} at the particle and field level using snapshot comparisons with the widely tested \texttt{2LPTic} code. The agreement in both real and redshift space power spectrum multipoles confirms that both the \texttt{CoLoRe‑2LPT} density and velocity fields correctly reproduce the expected mildly non‑linear evolution. We further showed that the interpolation of particle quantities onto a regular grid preserves clustering properties up to the mesh scale, making the resulting density and velocity fields suitable for tracer population and skewer extraction.

Second, we implemented a thresholded linear biasing scheme to populate the \texttt{CoLoRe-2LPT} lightcone density field with quasars (QSOs).
This biasing scheme is calibrated against \texttt{Abacus} Halo Occupation Distribution (HOD) QSO catalogues matching their clustering in real space up to $k\sim0.1$ $h$/Mpc within $1\sigma$. 
In redshift space, we additionally include a Gaussian random velocity component to model Fingers-of-God effects, maintaining agreement with \texttt{Abacus} down to scales comparable to those achieved in real space. The resulting QSO catalogues recover the latest DESI DR1 QSO large‑scale bias within $1-5\%$ for $z\in(1.8,3.8)$.

Then, skewers from each QSO to the observer are saved. \texttt{LyaCoLoRe-2LPT} is able to take these 2LPT density and velocity skewers and process them to get transmitted flux Ly$\alpha$ skewers with reliable small-scales to reproduce observational Ly$\alpha$ clustering statistics while keeping \texttt{CoLoRe-2LPT} large scale information. 

We generate 400 full-sky 10 Gpc/h \texttt{(Lya)CoLoRe-2LPT} mocks with $2.4$ Mpc/$h$ cell resolution and use 10 of them ($\sim$ 70 DESI realizations) to validate the performance of \texttt{LyaCoLoRe-2LPT} before adding any contaminant or DESI survey property. We measure the one dimensional power spectrum and the mean flux in these mocks getting results which are within 10 $\%$ of the DESI DR1 measurements for $z\in(1.8,3.8)$ and $k\lesssim0.01$ s/km. We also measure the Ly$\alpha$ auto-correlation and the Ly$\alpha\times\rm{QSO}$ cross-correlation, and perform BAO and FS fits on them using the DESI baseline model for $40 < r$ [Mpc/$h$] $<180$. These fits recover values for the Ly$\alpha$ bias and $\beta$ parameters within $\lesssim1\sigma$ of the DESI DR2 BAO measurement and effectively unbiased BAO and AP parameters for the statistical uncertainty of DESI DR2. Furthermore, as a signal of the better non-linearities of these mocks, we can reproduce directly the BAO broadening expected from theory in comparison to the ad-hoc BAO broadening introduced in past mocks. These non-linearities are also present, in contrast with previous Gaussian/log-normal mocks, when trying to push the linear model to smaller scales. We find that the linear model fit starts to fail and that the Arinyo non-linear model improves the fit for $r_{\rm{min}} < 40$ due to the stronger non-linearities in these mocks.

Finally, we generate 400 \textit{contaminated} mocks including the DESI DR2 survey characteristics, together with continuum fitting (and distorsion) and astrophysical contaminants: high column density systems (HCDs) and metals. Using a similar approach as with the QSO biasing, we populate the Ly$\alpha$ skewers with HCDs with $b_{\rm{HCD}}\sim2$ and an abundance ($dn/dz$) that matches the observations. For metals, we assume that their absorption is strictly linear to the Ly$\alpha$ one, tuning this proportionality to recover metal biases and shapes of the metal correlations, consistent to within $20 \%$ of the DESI DR2 correlation functions. The final validation of the pipeline consists of a complete BAO fit on these 400 \textit{contaminated} mocks. The equivalent fit for FS will be presented in the companion paper \citep{DESI:2026DR2IV}. This fit recovers parameters consistent with those of the \textit{raw} mocks and the DESI DR2 observational BAO fit, concluding the validation.

We conclude that the product mocks of this work are appropriate to be used in the validation process of BAO and FS projects, providing robust tests of analysis
pipelines and advancing the current state-of-the-art. The 400 mocks generated in this project have been key for the validation and modeling decisions of the DESI DR2 Ly$\alpha$ FS analysis presented in the companion papers \citep{DESI:2026DR2IV, MHVal}. We expect that, together with this FS analysis, these improved mocks will be used for future DESI -- and any up-coming survey -- BAO and FS analyses.

\section*{Data Availability}

Data from the plots in this paper will be made available on Zenodo (\url{https://zenodo.org/records/XXXXXX}) as part of DESI's Data Management Plan.

\begin{acknowledgements}
      MFRHB and SA were funded by the Agencia Estatal de Investigación (AEI, Spain, MCIN/AEI/10.13039/501100011033) and FSE+ (Europe) under projects PID2024-156844NA-C22 and PID2021-123012NB-C42. SA was further supported by the Ramon y Cajal  fellowship RYC2022-037311-I. AFR acknowledges financial support from the Spanish Ministry of Science and Innovation through the PID2024-159420NB-C41 project and the ``Excelencia Severo Ochoa'' program (CEX2024-001441-S from MICIU AEI 10.13039/501100011033) and the European Union through the ERC Consolidator Grant program (COSMO-LYA, grant agreement 101044612). IFAE is partially funded by the CERCA program of the Generalitat de Catalunya. HKHA acknowledges the funding of the French Agence Nationale de la Recherche (ANR) under grant ANR-22-CE31-0009 (HZ3DMAP project) and grant ANR-22-CE92-0037 (DESILya project). DA is funded by the Science and Technology Facilities Council under grant UKRI1164, and further acknowledges support from the Beecroft Trust. This material is based upon work supported by the U.S. Department of Energy (DOE), Office of Science, Office of High-Energy Physics, under Contract No. DE–AC02–05CH11231, and by the National Energy Research Scientific Computing Center, a DOE Office of Science User Facility under the same contract. Additional support for DESI was provided by the U.S. National Science Foundation (NSF), Division of Astronomical Sciences under Contract No. AST-0950945 to the NSF’s National Optical-Infrared Astronomy Research Laboratory; the Science and Technology Facilities Council of the United Kingdom; the Gordon and Betty Moore Foundation; the Heising-Simons Foundation; the French Alternative Energies and Atomic Energy Commission (CEA); the Secretariat of Science, Humanities, Technology and Innovation (SECIHTI) of Mexico; the Ministry of Science, Innovation and Universities of Spain (MICIU/AEI/10.13039/501100011033), and by the DESI Member Institutions: \url{https://www.desi.lbl.gov/collaborating-institutions}. Any opinions, findings, and conclusions or recommendations expressed in this material are those of the author(s) and do not necessarily reflect the views of the U. S. National Science Foundation, the U. S. Department of Energy, or any of the listed funding agencies.

The authors are honored to be permitted to conduct scientific research on I'oligam Du'ag (Kitt Peak), a mountain with particular significance to the Tohono O’odham Nation.

\end{acknowledgements}

\bibliographystyle{aa}
\bibliography{references_2lpt}

\appendix
\section{Validation of the HCD modelling}\label{AppendixB}
We validate the DLA modelling procedures of \autoref{sec:hcd} by computing the HCD $\times$ Ly$\alpha$ cross-correlation and measuring the incidence rate ($dn/dz$). 

\begin{figure}[h]
    \centering
    \includegraphics[width=\linewidth]{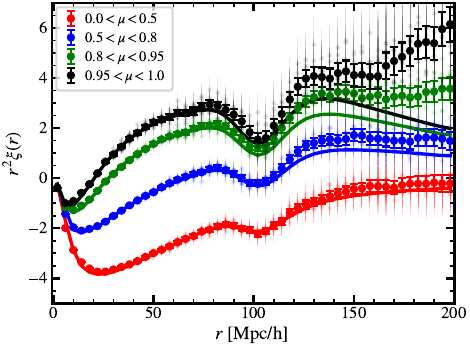}
    \caption{Ly$\alpha$ $\times$ DLA cross-correlation for the stack of 10 \textit{raw} \texttt{LyaCoLoRe-2LPT}. The solid lines represent the best BAO linear fit for $40 < r$ [Mpc/$h$] $ < 80$.}
    \label{fig:dlacross}
\end{figure}

We compute the HCD $\times$ Ly$\alpha$ cross-correlation following the same procedure as for the QSO cross-correlation in \autoref{sec:lya_auto} with the only caveat that we restrict ourselves to HCDs within $[1040, 1100]$ \AA{} following \citet{LyaCoLoRe} to avoid biases from the QSO $\times$ Ly$\alpha$ cross-correlation. This effect will be studied in more detail in a separate publication (Ruiz-Herrera Bernal et al., in prep). We then fit the HCD cross-correlation using the model described in \autoref{sec:lya_auto} fixing all the free parameters to the best fit values of \autoref{tab:fits} and leaving free the HCD $\beta$ parameter ($\beta_{\rm{HCD}}$) for $40 <r$ [Mpc/$h$] $< 80$. We present the cross-correlation measurement for the stack of 10 \textit{raw} \texttt{LyaCoLoRe-2LPT} mocks together with the best fit model in \autoref{fig:dlacross}. We can see that the model fits the data well for the used $r$ range, but starts failing at high $r$ due to the residual effect of the previously mentioned QSO $\times$ Ly$\alpha$ cross-correlation bias. We recover a best fit $\beta_{\rm{HCD}} = 0.4797 \pm 0.010$ which results in a HCD bias\footnote{We also performed the fit leaving free $b_{\rm{HCD}}$ instead of $\beta_{\rm{HCD}}$ obtaining equivalent results.} of $b_{\rm{HCD}} = 2.01 \pm 0.02$ consistent with the input value and the literature \citep{AndreuDLAs, IgnasiBiasDLAs, IgnasiBiasSBLAs}.

\begin{figure}[h]
    \centering
    \includegraphics[width=\linewidth]{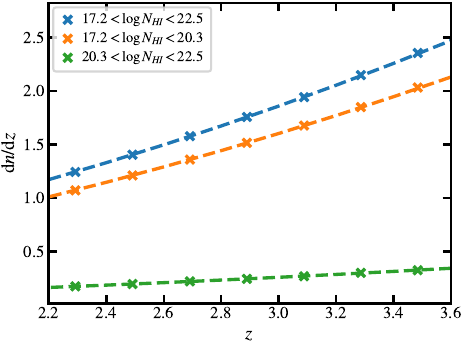}
    \caption{Comparison of the mean HCD $dn/dz$ in 10 \textit{raw} \texttt{LyaCoLoRe-2LPT} mocks (represented as crosses) with the \texttt{pyigm} \citep{ProchaskaDLA} model (represented as dashed lines) for the three main HCD column density bins over the DESI redshift range.}
    \label{fig:dndz_raw}
\end{figure}

Finally, we validate the incidence rate modelling by directly calculating $dn/dz$ in the previous 10 \textit{raw} \texttt{LyaCoLoRe-2LPT} mocks for the main types of HCD depending on their column density (see \autoref{sec:hcd}). We present the mean $dn/dz$ for these mocks together with the input \texttt{pyigm} \citep{ProchaskaDLA} model in \autoref{fig:dndz_raw}. We find consistency between the measured $dn/dz$ and the model for all redshifts and column density bins.

As an independent test, we also validated that the absorption we associate to the discrete HCDs with \texttt{quickquasars} (see Section \ref{sec:QQ}) has the expected clustering properties associated to the Rogers model  \citep{RogersMocks, eBOSSBAO} used in Section \ref{sec:val_cont}. We generate 10 HCD-only mocks, that is, mocks that only include the absorption associated to the HCD absorbers without Ly$\alpha$ forest transmission, measure the HCD (absorption) auto-correlation and the HCD (absorption) $\times$ QSO cross-correlation and fit them as described in Sections \ref{sec:lya_auto} and \ref{sec:val_cont}. We perform this fit on the stack of 10 mocks for $10 \leq r $ [Mpc/$h$] $\leq$ 80 and find $L_{\text{HCD}}=5.512 \pm 0.281$, $b_{\text{HCD(absorber)}} = -0.035 \pm 0.001$ and $\beta_{\text{HCD}}=0.485\pm0.035$. These values are completely consistent with the discrete clustering properties of the previous test and also with the DESI DR2 BAO best fit values (see \autoref{tab:bao_fits}).

\end{document}